\documentclass[a4paper]{article}
\pdfoutput=1 % if your are submitting a pdflatex (i.e. if you have
\usepackage{jcappub} % for details on the use of the package, please
\usepackage[T1]{fontenc} % if needed

\title{Test of the statistical isotropy of the universe using gravitational waves}

\author[a,b,1]{Giacomo Galloni,\note{Corresponding author.}}
\author[c,d,e]{Nicola Bartolo,}
\author[c,d,e,f]{Sabino Matarrese}
\author[a,b]{Marina Migliaccio}
\author[c,d]{Angelo Ricciardone}
\author[a,b]{and Nicola Vittorio}

\affiliation[a]{Dipartimento di Fisica, Università di Roma Tor Vergata,\\Via della Ricerca Scientifica, 1, 00133, Roma, Italy}
\affiliation[b]{INFN, Sezione di Roma 2,\\Via della Ricerca Scientifica, 1, 00133 Roma, Italy}
\affiliation[c]{Dipartimento di Fisica e Astronomia ``G. Galilei'', Universit\`a degli Studi di Padova,\\Via Marzolo 8, I-35131 Padova, Italy}
\affiliation[d]{INFN, Sezione di Padova,\\Via Marzolo 8, I-35131 Padova, Italy}
\affiliation[e]{INAF - Osservatorio Astronomico di Padova,\\Vicolo dell’Osservatorio 5, I-35122 Padova, Italy}
\affiliation[f]{Gran Sasso Science Institute,\\Viale F. Crispi 7, I-67100 L’Aquila, Italy}

\emailAdd{giacomo.galloni@roma2.infn.it}
\emailAdd{nicola.bartolo@pd.infn.it}
\emailAdd{sabino.matarrese@pd.infn.it}
\emailAdd{marina.migliaccio@roma2.infn.it}
\emailAdd{angelo.ricciardone@pd.infn.it}
\emailAdd{nicola.vittorio@uniroma2.it}

\abstract{Since WMAP and Planck some anomalous features appeared in the Cosmic Microwave Background (CMB) large-angle anisotropy, the so-called anomalies. One of these is the hemispherical power asymmetry, i.e. a difference in the average power on the two hemispheres centered around $(l,b) = (221, -20)$, which shows a relatively high level of significance. Such an anomaly could be the signature of a departure from statistical isotropy on large scales. Here we investigate the physical origin of this anomaly using the Cosmological Gravitational Wave Background (CGWB) detectable by future GW detectors. Indeed, the CGWB offers a unique window to explore the early universe and we show that it can be used in combination with CMB data to shed light on the statistical isotropy of our universe. Specifically, we study the evolution of gravitons in the presence of a modulating field in the scalar gravitational potentials accounting for the hemispherical power asymmetry and we infer the amplitude of this modulating field through a minimal variance estimator exploiting both constrained and unconstrained realizations of the CGWB. We show that the addition of the CGWB will allow an improvement in the assessment of the physical origin of the CMB power asymmetry. Accounting for the expected performances of LISA and BBO, we also show that the latter is expected to be signal-dominated on large-scales, proving that the CGWB could be the keystone to assess the significance of this anomaly.}
\begin{document}

\maketitle

\flushbottom

\section{Introduction}
The CMB power asymmetry has been first measured by the Wilkinson Microwave Anisotropy Probe (WMAP) satellite and then confirmed by the Planck mission. Both of them have shown an excess of power in a preferred hemisphere which seems to escape the unknown-systematics paradigm, showing a relatively high level of significance \cite{Eriksen_2004,Hansen_2009_WMAP,collaboration2019planck}. This can be a signature of a violation of statistical isotropy on large scales, that might have a cosmological origin, suggesting the presence of new physics. 
In the literature, people have tried to describe such an anomaly as an effect of a local break of isotropy (in our local observable Universe) due to a (dipole) modulation of the scalar gravitational potentials present in the Boltzmann equation for the CMB photons, which substantially involves the presence of perturbations on scales larger than the Hubble volume (e.g. during an early epoch of inflation). Such a modulation could stem in the presence of some non-Gaussianity, typically in the context of multiple-field inflationary scenarios \cite{Schmidt_2013,Byrnes_2015,Byrnes_2016,Ashoorioon_2016,Adhikari_2016,Hansen_2019}. These fields generate a local break of statistical isotropy through their gradient and act on the CMB \cite{Hu}, leaving as a relic the asymmetry we observe; all this without flawing the hypothesis of an underlying global isotropy of the Universe, since the modulation gets averaged over many Hubble volumes.
Aside the CMB, in the next future we expect to observe two other background signals in the Gravitational Wave (GW) sector. The first one is the so-called Astrophysical Gravitational Wave Background (AGWB), which comes from the overwhelming quantity of astrophysical sources that interferometers will not be able to detect separately and will be perceived as a stochastic background \cite{Regimbau:2011rp,Abbott_2019,Bertacca_2020}. On the other hand, we know that a smoking-gun of an inflationary period is a CGWB propagating from every direction in the sky as a direct consequence of the quantum nature of the metric fluctuations enhanced by inflation (see e.g. \cite{MAGGIORE2000283, Guzzetti:2016mkm, Bartolo_2016, Caprini_2015, Caprini_2018} for possible cosmological sources).
Besides their average contribution to the energy density of the Universe, these two backgrounds are also characterized by anisotropies (as the CMB, see e.g. \cite{Dodelson}), which are produced by their propagation through a perturbed Universe and at the time of their generation \cite{Alba_2016, Contaldi_2017, Geller:2018mwu, Bartolo_2019, Bartolo}. In spite of what happens for the CMB, where essentially anisotropies date back up to the Last Scattering Surface (LSS) \cite{peebles_1970,Bond_1983}, in the case of the CGWB they could provide us with crucial insights on the Early Universe, since the latter is transparent to GWs below the Planck energy scale \cite{Misner:1974qy,garoffolo_2020}.

Both backgrounds can be described using a Boltzmann approach. In particular, for GWs it is possible to define a distribution function for the gravitons, which will evolve according to the Boltzmann equation in a flat Friedmann-Lam\^aitre-Robertson-Walker (FLRW) Universe perturbed at first order in scalar and tensor perturbations as in \cite{Bartolo_2019, Bartolo}. The evolution will then depend on the gravitational potentials, i.e. the scalar first-order perturbations of the metric, which act like a ``source'' term for GWs.

In this work, we study the effect of a dipole modulation of these gravitational potentials (as predicted by models trying to explain CMB anomalies) on the CGWB anisotropies.
This primordial modulations can be plugged in the scalar sourced part of the Boltzmann equation, changing the evolution of the distribution function and its statistical properties, encoded, at the end of the day, in the angular power spectrum of the GW energy density.

In order to interface with observations, we extensively modified the Boltzmann code CLASS \cite{lesgourgues2011cosmic, class2} to compute the angular power spectrum of the CGWB to include the effects of a dipole modulation both for temperature anisotropies of the CMB and of the CGWB. On top of these we compute  the CGWBxCMB cross-correlation angular power spectrum. For both of them, we included the dependence on the tensor spectral tilt $n_t$ pointed out in \cite{Ricciardone_2021} in order to highlight how these spectra change in a reasonable range of $n_t$.
Subsequently, after having generalized some of the routines of the map-generation software HEALPix \cite{healpix} in order to account for the novel characteristics of the two observables in the presence of the modulation, we obtain constrained maps of the CGWB \cite{Bertschinger1987,Hoffman_1991,Bucher:2011nf, Kim:2012iq, Manzotti:2014kta}, where we exploit the information we already have from CMB high resolution measured sky maps.

Specifically, our statistical analysis can be divided in two parts: in the first we study how Cosmic Variance (CV) can ``simulate'' the presence of a modulation, without the need of any non-standard physics. For this we generate a set of 1000 unconstrained realizations of the CGWB and we study the Joint CV (JCV) of CMB$+$CGWB through an estimator defined in \cite{Hu}. In the second part, we estimate the amplitude of the modulating field from Planck's map and a set of 1000 constrained realizations of the CGWB, which still show the presence of a modulation. This indicates that if these models are indeed reproducing nature, we must see the same signal on the CGWB side. Thus, this can be a useful diagnostic tool for a future observation of this background signal (see \cite{Taruya_2005, Taruya_2006, baker2019high, Contaldi_2020_maps} for some insights on sky-maps of the GW backgrounds and the angular resolution of GW Astronomy).
This analysis is performed both including or not the instrumental noise coming from either LISA \cite{amaroseoane2017laser,Barausse_2020,Bartolo_2018_LISA,Caprini_2019,Flauger:2020qyi} or BBO \cite{BBO_2006}.

Summarizing some of the results of this work, consistently with \cite{valbusa_2020} the CGWB is affected both by the SW and ISW effect. In \cite{Hu}, treating the CMB case, the ISW was neglected, while in this work, focused on the CGWB, both effects are accounted for, including the primordial modulation of the gravitational potentials. As previously mentioned, we explore the dependency of the CGWB spectra and its cross-correlation with CMB on the tensor spectral tilt.
We prove that, analogously to the CMB, the CGWB acquires non-trivial couplings between multipoles. In other words, the correlation between multipoles is not diagonal anymore in $\ell$ (thus $\propto \delta_{\ell\ell^\prime}$), but a coupling between modes with $\ell$ to $\ell\pm1$ and $\ell\pm2$ appears. 
In addition, we find that the presence of the modulation produces a modification of the ISW contributions to the angular power spectrum of the CGWB, suppressing them in the off-diagonal terms of the correlation in a more prominent way w.r.t. the standard diagonal term.

Furthermore, in the noiseless case, we show that adding the CGWB to the joint estimation of the amplitude of the modulation of the gravitational potentials improves greatly the degree of significance of the power asymmetry. Specifically, the $83.4\%$ of distribution of the jointly estimated amplitudes coming from the 1000 CGWB realizations is above $3\sigma$-significance, whereas all of them improve the significance obtained using solely Planck's SMICA map without GWs, i.e. $1.95\sigma$.

Finally, in the noisy case, we show that BBO has the capacity to recover faithfully the noiseless expectations for JCV when we limit the range of multipoles to the first six and also is potentially able to assess the significance of the CMB power asymmetry, mostly for blue spectra. This is possible even sticking to a power-law description of the tensor power spectrum with values of the tilt compatible with the bounds of \cite{Planck_2018} when including LIGO-Virgo (i.e. $-0.76< n_t< 0.52$). 

The structure of this work is as follows: in section \ref{sec:boltzmann} we go through the Boltzmann approach to characterize both the CGWB and the CMB. Then, from section \ref{sec:modulation}, we treat the case in which we add a modulation in the Boltzmann equation (at the end of this section there are the main theoretical results of this work). In section \ref{sec:res_plot} we explore how the CGWB statistics changes in presence of the modulation, analyzing its angular power spectrum and correlation with CMB. In section \ref{sec:res_asses} we study the role of GWs in assessing the significance of the power asymmetry, both assuming instrumental noise or not. In section \ref{sec:conclusions}, we summarize the results of this paper.
%__________________________________________________________________

\section{Boltzmann approach for GWs and CMB} \label{sec:boltzmann}
Exploiting the statistical nature of primordial GWs, we can define a distribution function for gravitons as $f_{GW}=f_{GW}(x^\mu,p^\mu)$, which in general depends on their position $x^\mu$ and momentum along their trajectory $p^\mu(x)$. This function will evolve according to the well-known Boltzmann equation, in the very same way of the CMB photons distribution $f_\gamma(x^\mu,p^\mu)$ (see e.g. \cite{Bartolo_2006, Bartolo_2007}). The gravitons distribution can be expanded as
\begin{gather}
     f_{GW}(\eta, \Vec{x}, \Vec{q}) = \Bar{f}_{GW}(q) -q\pdv{\Bar{f}_{GW}}{q}\Gamma(\eta, \Vec{x}, \Vec{q})\ , 
\end{gather}
where $\eta$ is the conformal time, $q=a\abs*{\Vec{p}}$ is the comoving momentum modulus, $\Gamma$ is a rescaling function of the first order term, which simplifies the subsequent equations \cite{Contaldi_2017} and $\Bar{f}_{GW}$ is the homogeneous and isotropic solutions of the Boltzmann equation.

Just like any other energy source of the Universe, GWs contribute to the overall energy density budget. The rescaling function $\Gamma$ is then related to the GWs energy contrast $\delta_{GW}$ and to the average contribution to the energy density $\ols{\Omega}_{{GW}}$ through \cite{Bartolo_2019, Bartolo_2020}
\begin{equation}
    \delta_{GW} (\eta_0,\Vec{x},q,\hat{n}) \equiv \qty[4 - \pdv{\ln \ols{\Omega}_{GW}}{\ln q}]\Gamma(\eta_0,\Vec{x},q,\hat{n})\ ,
\label{eq:energy_contrast}
\end{equation}
where $\eta_0$ is the conformal time today and $\hat{n}$ is the GW direction of motion.  
In this work we consider a power-law description of $\ols{\Omega}_{GW}(q) \propto q^{n_t}$, thus the density contrast and $\Gamma$ are simply related through $\delta_{GW}=(4-n_t)\Gamma$ \cite{Ricciardone_2021}.
We consider a flat FLRW metric perturbed at linear level,  in the Poisson's gauge. This results in having 
\begin{equation}
    ds^2=a^2(\eta)\qty[-(1+2\Phi)d\eta^2 + \left[\left(1-2\Psi\right)\delta_{i j}+\chi_{i j}\right]dx^idx^j]\ ,
\label{eq:flrw_metric}
\end{equation}
where $a(\eta)$ is the scale factor, $\Phi,\Psi$ are the scalar perturbations called gravitational potentials and $\chi_{ij}$ is the Transverse-Traceless (TT) tensor perturbations, for which $\chi_i^{i}=0=\partial^i\chi_{ij}$.

Thus, we can write the first order Boltzmann equation for the CGWB as
\begin{equation}
    \pdv{\Gamma}{\eta} + n^i\pdv{\Gamma}{x^i} = \frac{d\Psi}{d\eta} - \frac{d\Phi}{dx^i} n^i - \frac{1}{2}\frac{d\chi_{jk}}{d\eta}n^jn^k\ .
\label{eq:real_space_source}
\end{equation}
The source function on the right hand side contains the scalar and tensor perturbations, which will affect the propagation of gravitons, providing the anisotropies of their distribution. 

For the CMB, the equations would have been nearly the same, however it is useful to underline a difference. CMB photons were thermally coupled before the recombination epoch, thus the continuous and very efficient scatterings erased any trace of the initial conditions, leaving behind a nearly ``memory-less'' thermal plasma. Once they decoupled, they started the free-streaming. Gravitons instead have never being thermal (at least below the Planck energy \cite{Misner:1974qy}), thus $\Gamma$ retains in general a $\order{1}$ dependency on the frequency, carrying information on the production mechanism \cite{Bartolo_2019, Ricciardone_2021}.

We do not report here the tensor source term $\Gamma_T$, which is computed as in \cite{Bartolo_2019,Bartolo} (see also appendix \ref{app:SvsT} for more details); thus, we can write the expression of the first order solution induced by the scalar perturbations after moving to Fourier space and decomposing the solution in spherical harmonics \cite{Bartolo_2019,Bartolo}:
\begin{equation}\begin{aligned}
    \Gamma_{\ell m,S} (\eta_0) = & 4\pi (-i)^\ell\int\frac{d^3k}{(2\pi)^3} Y^*_{\ell m}(\hat{k}) \int_{\eta_{\text{in}}}^{\eta_0} d\eta \Bigg\{ \delta(\eta-\eta_{in}) \qty[\Phi(\eta,\Vec{k}) + \Gamma_I(\eta,\Vec{k},q)]  \\
    & + \frac{\partial\qty[\Psi(\eta,\Vec{k}) + \Phi(\eta,\Vec{k})]}{\partial\eta}  \Bigg\}j_\ell\qty[k\qty(\eta_0-\eta)]\ ,
\label{eq:scalar_term_gamma}
\end{aligned}\end{equation}
where $_S$ reminds us that we are looking at scalar sourced anisotropies and $\Gamma_I(\eta_{in},\Vec{k},q) = -2/(4-n_t)\Phi(\eta_{in},\Vec{k})$ is a term coming from the initial conditions \cite{Ricciardone_2021}.
In Eq.\ref{eq:scalar_term_gamma}, the first term represents the SW effect, i.e., an anisotropy set by the value of the gravitational potential $\Phi$ at $\eta_{\text{in}}$. The second one accounts for the propagation of gravitons from $\eta_{in}$ up to us today, thus the ISW effect.
Interestingly, this procedure is similar to the CMB one, for which, including only the SW and ISW effects, we have \cite{Dodelson}
\begin{equation}\begin{aligned}
    \Theta_{\ell m,S}(\eta_0) = & 4\pi (-i)^\ell\int\frac{d^3k}{(2\pi)^3} Y^*_{\ell m}(\hat{k}) \int_{\eta_{\text{in}}}^{\eta_0} d\eta \Bigg\{g(\eta) \qty(\Theta_0(\eta, \Vec{k}) + \Phi(\eta,\Vec{k}) )  \\
    & + e^{-\tau(\eta)}\frac{\partial\qty[\Psi(\eta,\Vec{k}) + \Phi(\eta,\Vec{k})]}{\partial\eta} \Bigg\} j_\ell\qty[k\qty(\eta_0-\eta)]\ .
\label{eq:scalar_term_gamma_cmb}
\end{aligned}\end{equation}
where $\tau=\int_\eta^{\eta_0} d\eta^\prime n_e \sigma_T a$ is the optical depth, $\sigma_T$ is the Compton cross section and $g(\eta) = -\tau^\prime(\eta)e^{-\tau(\eta)}$ is the visibility function.

Let us remark again a very important detail: notice that in Eq.\ref{eq:scalar_term_gamma} $\Phi$ and $\Gamma_I$ are evaluated in $\eta_{in}$, which roughly corresponds to the end of inflation, instead in Eq.\ref{eq:scalar_term_gamma_cmb} there is the visibility function, which restricts the evaluation of $\Phi$ and $\Theta_0$ around recombination. This is the consequence of the fact that CMB photons are generated approximately on the LSS, whereas we consider the primordial gravitons to be produced all together at the end of inflation.
In order to relate these quantities to some observables, we compute the angular power spectrum, i.e the two point correlation function assuming that the statistical variables involved satisfy statistical isotropy. In this case, we can write
\begin{gather}
    \expval{\Gamma_{\ell m}(\eta_0)\Gamma_{\ell^\prime m^\prime}^*(\eta_0)} = \delta_{\ell \ell^\prime} \delta_{m m^\prime} C_{\ell}^{CGWB} \qq{,}
    \expval{\Theta_{\ell m}(\eta_0)\Theta_{\ell^\prime m^\prime}^*(\eta_0)} = \delta_{\ell \ell^\prime} \delta_{m m^\prime} C_{\ell}^{CMB} \ .
\label{eq:std_correlators}
\end{gather}

%----------------------------------------------------------------------------------------

\section{Modulation of the gravitational potentials} \label{sec:modulation}
What happens if we want to include a modulation of the gravitational potentials, relaxing some of the assumption of the standard case?

The scalar perturbations of the metric are typically expressed as 
\begin{equation}
    \Phi = \zeta(\Vec{k})\times\qty{\text{Transfer Function}(k)}\times\qty{\text{Growth Function}(\eta)}\ ,
\label{eq:pert_decomp}
\end{equation}
where $\zeta(\Vec{k})$ is the primordial value of the curvature perturbation set during inflation, the \{Transfer Function(k)\} controls the evolution of perturbations through the epochs of horizon crossing and radiation/matter transition and the \{Growth Function($\eta$)\} controls the wavelength-independent growth at late times \cite{Dodelson}.

From now on, except where differently specified, we will refer with the name ``transfer function'' to the actual product of \{Transfer Function$(k)$ $\times$ Growth Function$(\eta)$\} of Eq.\ref{eq:pert_decomp}, in such a way that we can write
\begin{equation}
    \Phi(\eta, \Vec{k}) = T_{\Phi}(\eta, k)\zeta(\Vec{k}) \qq{ , } \Psi(\eta, \Vec{k}) = T_{\Psi}(\eta, k)\zeta(\Vec{k})\ . 
\label{eq:gravi_potentials}
\end{equation}
Neglecting any anisotropic stress, we could consider for simplicity $T_\Phi=T_\Psi$, but we keep them distinguished to allow future generalization of this work. 

In order to include a modulation of the gravitational potentials, we can assume that $\Phi(\vec{x})$ actually depends on two fields $g(\Vec{x})$ and $h(\Vec{x})$, instead of a single field $\zeta$; $h$ is only related to super-horizon scale fluctuations, carrying the modulation, and $g$ has only sub-horizon modes \cite{Hu}. In this way, $h$ will assume a deterministic value in our Hubble volume, while the other field will look like standard stochastic fluctuations and will account for the standard Gaussian behavior of the anisotropies. As aforementioned, across the Hubble volume an observer will see spontaneously broken statistical isotropy as an effect of the slow modulation of $h$, while its local gradient and curvature will pick a preferred direction, breaking the statistical isotropy \cite{Byrnes_2015}. Specifically, we write $\zeta$ as
\begin{equation}
    \zeta(\Vec{x}) = g(\Vec{x})[1+h(\Vec{x})]\ ,
\end{equation}
where again $g(\Vec{x})$ makes our considerations compatible with the observed statistical homogeneity and isotropy on small scales and $h(\Vec{x})$ is the modulating field breaking isotropy.
Going to Fourier space, we will perform the ensemble averages on one Hubble volume, so that only the $g$ field will get averaged. In this way we find
\begin{gather}
    \zeta(\Vec{k}) = g(\Vec{k}) + \int\frac{d^3k^\prime}{(2\pi)^3} g(\Vec{k}^\prime) h(\Vec{k} - \Vec{k}^\prime)
\end{gather}
and the 2-point correlation function of $\zeta$ becomes
\begin{equation} \begin{aligned}
\left\langle\zeta(\Vec{k}) \zeta^{*}\left(\Vec{k}^{\prime}\right)\right\rangle=&(2 \pi)^{3} \delta\left(\Vec{k}-\Vec{k}^{\prime}\right) P_{g}(k) +\left[P_{g}(k)+P_{g}\left(k^{\prime}\right)\right] h\left(\Vec{k}-\Vec{k}^{\prime}\right) \\
&+\int \frac{d^{3} \widetilde{k}}{(2 \pi)^{3}} P_{g}(\widetilde{k}) h(\Vec{k}-\widetilde{\Vec{k}}) h^{*}\left(\Vec{k}^{\prime}-\widetilde{\Vec{k}}\right)\ ,
\label{eq:phicorrelation}
\end{aligned}\end{equation}
where $P_{g}(k)$ is the primordial power spectrum of scalar perturbations, characterized as a power-law:
\begin{equation}
    P_{g}(k) = A_s \qty(\frac{k}{k_*})^{n_s-1}\ ,
\end{equation}
where $A_s$ is the amplitude at an arbitrary pivot scale $k_* = 0.05$ Mpc$^{-1}$ and $n_s$ is the scalar spectral index.
In Eq.\ref{eq:phicorrelation} we can identify three terms, based on their order in $h(\Vec{k})$.

Recalling that we are looking to a modulation of the scalar perturbations, the interesting term to explore is $\Gamma_S$, thus we can plug the previous equation into
\begin{equation} \begin{aligned}
    \left\langle\Gamma_{\ell m, S}\Gamma^{*}_{\ell^{\prime} m^{\prime}, S} \right\rangle=&(4 \pi)^{2}(-i)^{\ell-\ell^{\prime}} \int \frac{d^{3} k}{(2 \pi)^{3}} \mathrm{e}^{i \vec{k} \cdot \vec{x}_{0}} \int \frac{d^{3} k^{\prime}}{(2 \pi)^{3}} \mathrm{e}^{-i \vec{k}^{\prime} \cdot \vec{x}_{0}} \left\langle\zeta(\Vec{k}) \zeta^{*}\left(\Vec{k}^{\prime}\right)\right\rangle Y_{\ell m}^{*}(\hat{k}) Y_{\ell^{\prime} m^{\prime}}\left(\hat{k}^{\prime}\right) \\ 
    & \times \ensuremath{\varT}_{\ell}^{S}\left(k, \eta_{0}, \eta_{\mathrm{in}}\right)\ensuremath{\varT}_{\ell^\prime}^{S}\left(k^\prime, \eta_{0}, \eta_{\mathrm{in}}\right)\ ,
\end{aligned}\end{equation}
where the transfer function is defined as
\begin{equation}\begin{aligned}
    & \ensuremath{\varT}_{\ell}^{S}(k, \eta_{0}, \eta_{in}) \equiv T_{\Phi}\left(\eta_{\text {in }}, k\right) j_{\ell}\left(k\left(\eta_{0}-\eta_{\text {in }}\right)\right) + \int_{\eta_{\text {in}}}^{\eta_{0}} d \eta^{\prime} \frac{\partial\qty[T_{\Psi}\left(\eta^{\prime}, k\right)+T_{\Phi}\left(\eta^{\prime}, k\right)]}{\partial \eta^{\prime}} j_{\ell}\left(k\left(\eta_{0}-\eta^{\prime}\right)\right)\ .
\end{aligned}\end{equation}

So, we obtain three different terms contributing to the two-point correlation functions:
\begin{equation}\begin{aligned}
     \expval{\Gamma_{\ell m, S}\Gamma^{*}_{\ell^{\prime} m^{\prime}, S}} = & \expval{\Gamma_{\ell m, S}\Gamma^{*}_{\ell^{\prime} m^{\prime}, S}} ^{(0)} + \expval{\Gamma_{\ell m, S}\Gamma^{*}_{\ell^{\prime} m^{\prime}, S}} ^{(1)} + \expval{\Gamma_{\ell m, S}\Gamma^{*}_{\ell^{\prime} m^{\prime}, S}} ^{(2)}\ ,
\label{eq:total_correlator}
\end{aligned}\end{equation}
where $(0, 1, 2)$ indicates the order in the modulating field. 
Let us now report the final results for each of these terms.

\subsection{Zeroth order term in the modulating field}
Starting from the zeroth order term (in the modulating field), it is easy to find
\begin{equation}
\left\langle\Gamma_{\ell m, S} \Gamma^{*}_{\ell^{\prime} m^{\prime}, S}\right\rangle ^{(0)}= \delta_{\ell \ell'} \delta_{mm'} \mathcal{C}_\ell^{(0)} \ ,
\end{equation}
where the angular power spectrum reads
\begin{equation}
  \mathcal{C}_\ell^{(0)} \equiv 4\pi \int \frac{d k}{k} \frac{k^3}{2\pi^2}P_{g}(k) \ensuremath{\varT}_{\ell}^{(S)2}\left(k\right)\ .
\label{eq:zeroth}
\end{equation}
As expected, this expression is basically the same obtained in the isotropic case.

\subsection{First order term in the modulating field}
To proceed further with the first order term, we need to specify an expression for the modulating field.

In our case, we want to reproduce an excess power in one of the two hemispheres of the CMB, thus the most natural and simple choice is to go for a dipole modulation following \cite{Hu}, such as 
\begin{equation}
    h(\Vec{x}) =\omega_1\sqrt{\frac{3}{4\pi}} \frac{1}{k_0D_{\text{rec}}}\sin{\Vec{k}_0\cdot\Vec{x}}\ ,
\label{eq:modulating_field}
\end{equation}
\begin{equation}
h(\Vec{k})=\frac{\omega_1}{2 i} \sqrt{\frac{3}{4 \pi}} \frac{(2 \pi)^{3}}{k_{0} D_{\mathrm{rec}}}\left[\delta\left(\Vec{k}-\Vec{k}_{0}\right)-\delta\left(\Vec{k}+\Vec{k}_{0}\right) \right]\ .
\end{equation}
where $\Vec{k}_0$ is the wavenumber of the modulating field fluctuation, $\omega_1$ is the amplitude of the modulation, with the subscript ${}_1$ reminding us that we are considering a dipole modulation and $D_{\mathrm{rec}}$ is the conformal distance to the LSS.

Here we notice some peculiarity of Eq.\ref{eq:phicorrelation}: working out the expression for $h\qty(\Vec{k}-\Vec{k}^{\prime})$, one can see that not only the modes with $\Vec{k}=\Vec{k}^{\prime}$ get correlated through the Dirac's delta, but also modes with $\Vec{k}^{\prime}=\Vec{k} \pm \Vec{k}_0$.
As a first approximation we can think the modulation in Eq.\ref{eq:modulating_field} as $\propto Y_{10}$ (dipole), but, again, one could have considered something different. For instance, \cite{Hu} considers also a quadrupolar modulation $\propto Y_{20}$ to study another CMB anomaly, i.e. the alignment of the quadrupole and octopole.

At the end, the computation of the first order term of Eq.\ref{eq:total_correlator} yields
\begin{equation}
\left\langle\Gamma_{\ell m, S} \Gamma^{*}_{\ell^{\prime} m^{\prime}, S}\right\rangle ^{(1)} = \omega_1 \delta_{mm^\prime} \qty[R_{\ell^\prime m}^{1, \ell}\mathcal{C}_\ell^{(1)} + R_{\ell m}^{1, \ell^\prime}\mathcal{C}_{\ell^\prime}^{(1)}]\ ,
\label{eq:first_order_term}
\end{equation}
where \footnote{Here we have implicitly approximated $\left[P_{g}(k)+P_{g}\left(\abs*{\Vec{k}+\Vec{k}_0}\right)\right] \simeq 2 P_{g}(k)$, given that we are working in the assumption of $k_0 \ll k$ since the modulating field $h(\Vec{x})$ has only super-horizon modes.}
\begin{equation}
    \mathcal{C}_\ell^{(1)} \equiv 4\pi \int \frac{dk}{k} \frac{k^3}{2\pi^2}  P_{g}(k) \ensuremath{\varT}_{\ell}^{S}\left(k, \eta_{0}, \eta_{\mathrm{in}}\right) \times \varU^\star_\ell\qty(\abs*{\Vec{k}+\Vec{k}_0}) \ .
    \label{eq:first_Cell}
\end{equation}
Here, we define two new transfer functions
\begin{gather}
    \varU^\star_\ell\qty(\abs*{\Vec{k}+\Vec{k}_0}) \equiv \ensuremath{\varT}_{\ell}^{SW}\qty(\abs*{\Vec{k}+\Vec{k}_0})+ \ensuremath{\varT}_{\ell}^{ISW\star}\qty(\abs*{\Vec{k}+\Vec{k}_0}) \ ,
\end{gather}
which are different from the regular $\ensuremath{\varT}_{\ell}^{S}\left(k, \eta_{0}, \eta_{\mathrm{in}}\right)$ since $T_\Phi,\ T_\Psi$ are evaluated in $\abs*{\Vec{k}+\Vec{k}_0}$ and the ISW$\star$ term contains the extra factor $\frac{\eta_0-\eta}{\eta_0-\eta_{in}}$: 
\begin{equation}
    \label{eq:transfer_1}
    \ensuremath{\varT}_{\ell}^{SW}\qty(\abs*{\Vec{k}+\Vec{k}_0}) \equiv T_{\Phi}\left(\eta_{\text {in }}, \abs*{\Vec{k}+\Vec{k}_0}\right) j_{\ell}(k \left(\eta_{0}-\eta_{\text {in}}\right))\ ,
\end{equation}
\begin{equation}\begin{aligned}
\label{eq:transfer_3}
    & \ensuremath{\varT}_{\ell}^{ISW\star}\qty(\abs*{\Vec{k}+\Vec{k}_0}) \equiv \int_{\eta_{\text {in }}}^{\eta_{0}} d \eta^{\prime} 
    \frac{\partial\left[T_{\Psi}\qty(\eta^{\prime}, \abs*{\Vec{k}+\Vec{k}_0})+T_{\Phi}\qty(\eta^{\prime}, \abs*{\Vec{k}+\Vec{k}_0})\right]}{\partial \eta^{\prime}} \frac{\eta_0-\eta^\prime}{\eta_{0}-\eta_{in}}
    j_{\ell}\left(k\left(\eta_{0}-\eta^{\prime}\right)\right)\ .
\end{aligned}\end{equation}
In Eq.\ref{eq:first_order_term}, the following coupling matrix appears
\begin{equation}\begin{aligned}
R_{\ell m}^{\ell_{1}, \ell_{2}} \equiv \,&(-1)^{m} \sqrt{\frac{(2 \ell+1)\left(2 \ell_{1}+1\right)\left(2 \ell_{2}+1\right)}{4 \pi}} \left(\begin{array}{ccc}
\ell_{1} & \ell_{2} & \ell \\
0 & 0 & 0
\end{array}\right)\left(\begin{array}{ccc}
\ell_{1} & \ell_{2} & \ell \\
0 & m & -m
\end{array}\right)\ ,
\end{aligned}\end{equation}
which couples in our case (i.e. $\ell_1 = 1$ and $\ell_2 = \ell^\prime$) modes with $\ell$ to $\ell \pm 1$ through the triangle rule of the $3-j$ Wigner's symbols \cite{Komatsu}.
\subsection{Second order term in the modulating field}
For what regards the second order term in the modulating field, we obtain \footnote{\label{note:k_approx}We implicitly assume here that the transfer function in $\widetilde{k}\pm k_0$ is approximately equal to the one evaluated in $\widetilde{k}$.}:
\begin{equation}
    \left\langle\Gamma_{\ell m, S} \Gamma^{*}_{\ell^{\prime} m^{\prime}, S}\right\rangle ^{(2)}  =  \omega_1^2 \delta_{mm^\prime}  \sum_j R_{\ell m}^{1, j}R_{\ell^\prime m}^{1, j} \mathcal{C}_j^{(2)}\ ,
\label{eq:second_order_term}
\end{equation}
where
\begin{equation}
     \mathcal{C}_\ell^{(2)} \equiv 4\pi \int \frac{d k}{k} \frac{k^3}{2\pi^2} P_{g}(k)\varU_{\ell}^{\star}(k) \varU_{\ell}^{\star}(k)\ .
     \label{eq:second_Cell}
\end{equation}

In Eq.\ref{eq:second_order_term}, the particular configuration of coupling matrices R couples $\ell$ to $\ell \pm 2$ and also gives a contribution to the diagonal part with $\ell = \ell^\prime$. 

Eq.\ref{eq:first_order_term} to Eq.\ref{eq:second_Cell} (with the exception of the definition of the coupling matrices \cite{Hu}) are the main theoretical results of this work.

With this set of equations we can already say that a dipole gets mainly excited by the first order term through the off-diagonal elements with $\ell = 2, \ell^\prime = 1$, or  $\ell = 1, \ell^\prime = 2$, even if it gets a small contribution from the second order one with $\ell = \ell^\prime = 1$.

Also, one can notice that approximating these results for $\Vec{k} +\Vec{k}_0 \simeq \Vec{k}$ and neglecting the ISW effect, we get $\varU_\ell^\star \simeq \ensuremath{\varT}_\ell^S$, thus $\ensuremath{\mathcal C}_\ell^{(0)} \simeq \ensuremath{\mathcal C}_\ell^{(1)} \simeq \ensuremath{\mathcal C}_\ell^{(2)}$. In other words, the results obtained in \cite{Hu} are reproduced by our computation. In particular they assumed that $T_\Phi = T_\Psi = -\frac{1}{3}$ to simplify the computations.

One can realize that the main difference between the CMB and CGWB is the specific shape of their ``source functions'', so Eq.\ref{eq:scalar_term_gamma} and Eq.\ref{eq:scalar_term_gamma_cmb}. The rest of the treatment is quite similar and this allows to exploit our results also for the CMB SW and ISW effects.

\section{Statistics in the presence of a dipolar modulation} \label{sec:res_plot}
In order to compute the contributions $\ensuremath{\mathcal C}_\ell^{(i)}$ of Eq.\ref{eq:first_Cell} and Eq.\ref{eq:second_Cell} we modified the publicly available Boltzmann code CLASS \cite{lesgourgues2011cosmic, class2}, both for the CGWB and the CMB. We add in the code the CGWB similarly to \cite{valbusa_2020} by mimicking the CMB procedures, while modifying the source functions from Eq.\ref{eq:scalar_term_gamma_cmb} to Eq.\ref{eq:scalar_term_gamma}. Also the tensor induced contribution is reproduced by mimicking the CMB procedure and exploiting the equations of \cite{Bartolo_2019,Bartolo} (see appendix \ref{app:SvsT} for the explicit expressions of the source functions). Then, we modify the vector of values of the mode $k$ where the gravitational potential transfer functions are evaluated throughout the code in order to shift them to $k + k_0$ (remember Eq.\ref{eq:transfer_1} and Eq.\ref{eq:transfer_3}). Furthermore, we compute the cross-correlation between the CMB and the CGWB, which will be crucial to compare the two observables \cite{Adshead:2020bji, Malhotra:2020ket, Ricciardone_2021, Dimastrogiovanni:2021mfs}.

Let us recall that in \cite{Hu} the ISW term was neglected, whereas in our modification of CLASS we implement it. Moreover, in our computation we keep the full transfer function expression encoded in CLASS, without assuming any approximation.
Also, remember that we assumed $k \ll k_0$ to obtain the expression of $\mathcal{C}_\ell^{(1)}$ and $\mathcal{C}_\ell^{(2)}$, thus the results should only be trusted when that condition is satisfied.

\begin{figure}
    \centering
    \includegraphics[width=.49\hsize]{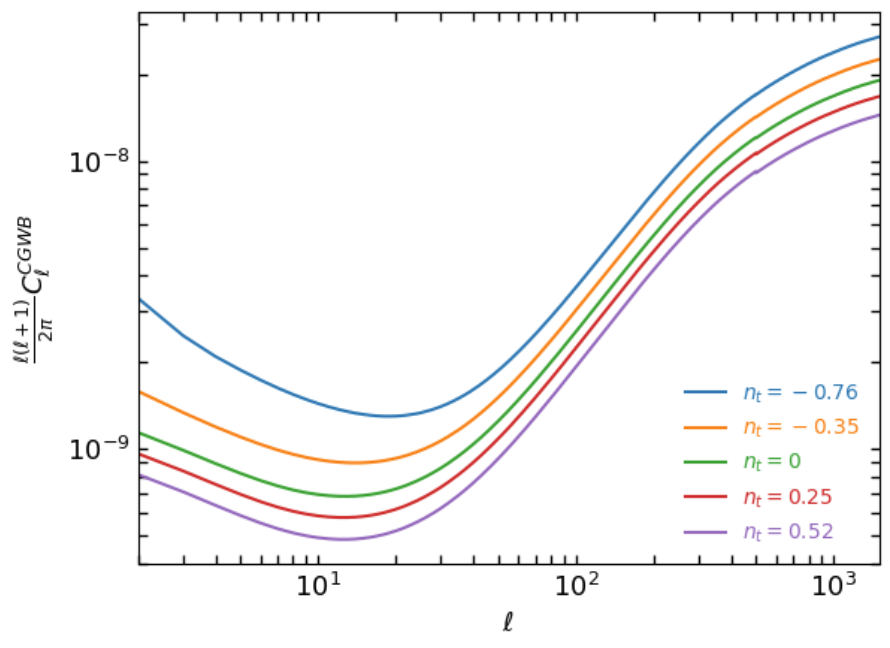}
    \hfill
    \includegraphics[width=.49\hsize]{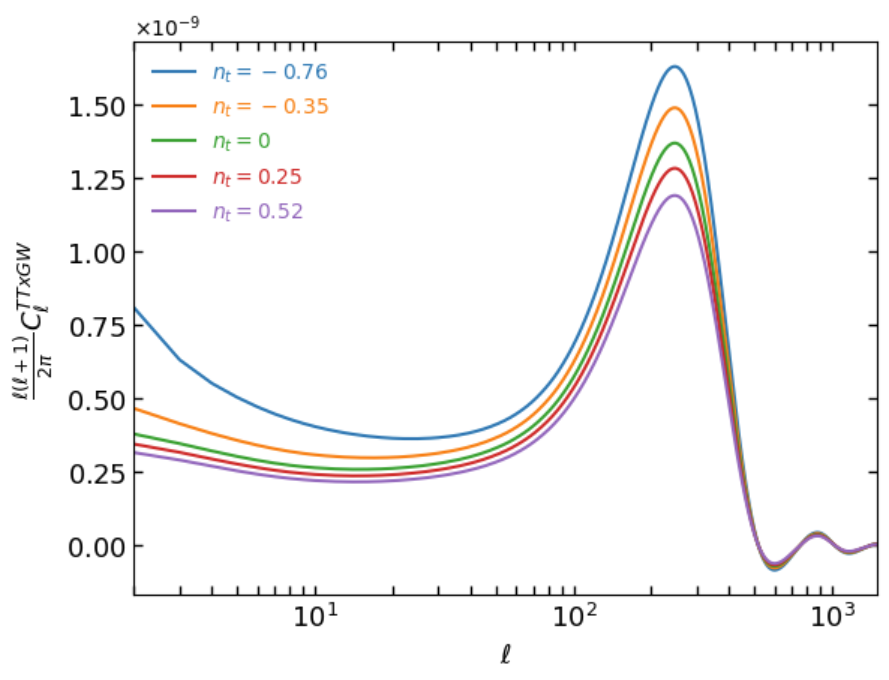}
    \caption{CGWB angular power spectrum (left) and cross-correlation angular power spectrum between CMB temperature and CGWB (right) when we assume different values for the tensor spectral tilt $n_t$ and we do not consider any effect due to the modulation ($\omega_1=0$).}
    \label{fig:CGWB_nt}
\end{figure}

Before looking at the effects of the modulation, it is worth exploring how the tensor spectral tilt affects the theoretical angular power spectrum of the CGWB following Eq.\ref{eq:scalar_term_gamma}. Figure \ref{fig:CGWB_nt} show respectively the CGWB angular power spectrum (left) and its cross correlation spectrum with CMB temperature anisotropy (right) (in appendix \ref{app:SvsT} we distinguish the scalar and tensor contributions to the spectra). 
These two figures need a further clarification: when we look at the CGWB, the actual observable we can measure with interferometers is the GWs energy density and its anisotropies, i.e. the energy contrast as a function of the direction in the sky. As aforementioned, this is related to the anisotropies of $\Gamma$ through Eq.\ref{eq:energy_contrast}. Thus, be aware that from this point on we will work with the angular power spectrum of the density contrast, gaining extra $(4-n_t)$-factors when we compute the CGWB, or its cross-correlation with the CMB temperature. For this reason, these two will depend on the spectral tilt $n_t$ both through the $\Gamma\to\delta_{GW}$ conversion factor and through the presence of $\Gamma_I$ in Eq.\ref{eq:scalar_term_gamma}. However, this is not all: Eq.\ref{eq:real_space_source} shows that there is another term contributing to the CGWB anisotropies induced by the presence of tensor perturbations of the metric $\chi_{ij}$ in Eq.\ref{eq:flrw_metric}. The transfer function of this term is proportional to the time derivative of the tensor mode function when decomposing the perturbation in circular polarizations (see e.g. \cite{Bartolo_2018_LISA}), i.e. $\propto -\chi^\prime(\eta,k)$. Then, the angular power spectrum will depend on the power-law description of the primordial tensor power spectrum \cite{Bartolo_2019}. In the case of CMB \cite{Seljak_1996}, the transfer function relative to tensor perturbations is also $\propto -\chi^\prime$, thus red tilted spectra are not only expected to enhance both the tensor contributions of the angular power spectra of CGWB and TT CMB at large scales, but also their cross-correlation (see appendix \ref{app:SvsT} for some further detail).

In order to fully explore the dependence on $n_t$, we choose some values for the tilt that span the bounds provided by \cite{Planck_2018} when including the data from LIGO-Virgo, i.e. $-0.76< n_t< 0.52$. Specifically, we choose $n_t = \qty{-0.76,\ -0.35,\ 0,\ 0.25,\ 0.52}$. It should be emphasized once again that we are assuming a power-law description of the tensor power spectrum in an inflationary context. Also, from this point on, all the six $\Lambda$CDM parameters are set to the best fit of \cite{Planck_parameters}, whereas the tensor-to-scalar ratio is chosen to saturate the bound provided by \cite{Planck_2018} when including the data from LIGO-Virgo, i.e. $r=0.066$ at $0.01$ Mpc$^{-1}$.

Going back to the CGWB angular power spectrum, the left panel of figure \ref{fig:CGWB_nt} shows how the spectrum changes when we assume different values of $n_t$. Recalling Eq.\ref{eq:scalar_term_gamma} and reminding the presence of the tensor sourced contribution, one can see that the higher the tilt, the more the anisotropies will be suppressed.
This same effect is brought to the cross-spectrum, as shown in the right panel of figure \ref{fig:CGWB_nt}. Interestingly, one can notice that for very red tilted spectra, the total cross-spectrum at large scales tends to increase significantly due to tensor contribution, as expected from the previous discussion of the tensor sourced contribution (see also appendix \ref{app:SvsT}).

\subsection{Contributions to the correlators}
Let us remind that $\mathcal{C}_\ell^{(1)}$ and $\mathcal{C}_\ell^{(2)}$ are just a part of the whole expression of the two novel contributions to the correlators (see Eq.\ref{eq:first_order_term} and Eq.\ref{eq:second_order_term}), thus more work has to be done to include the proportionality coefficients.
The following step to simplify the computation would be to fix the value of $m$, given that the correlators are ``diagonal'' in $m$, and compute Eq.\ref{eq:first_order_term} and Eq.\ref{eq:second_order_term}. 
Using HEALPix \cite{healpix}, we are able to compute a realization of the $\Theta_{\ell m}$ once the angular power spectrum is specified. In the standard case the $\Theta_{\ell m}$ are distributed around a Gaussian with zero mean and a variance given by the angular power spectrum \footnote{In the standard case there is no correlation between multipoles, thus the variance of the $\Theta_{\ell m}$ is the same for every $m$ once you fix $\ell$ and is solely controlled by the value of $\ensuremath{\mathcal C}_\ell$.}.
So, we modified some of the routines of HEALPix in order to account for the correlation between multipoles introduced by the modulation, but, most importantly, for the fact that now the results will depend on $m$ (i.e. broken statistical isotropy).

\begin{figure}
    \centering
    \includegraphics[width=.4\hsize]{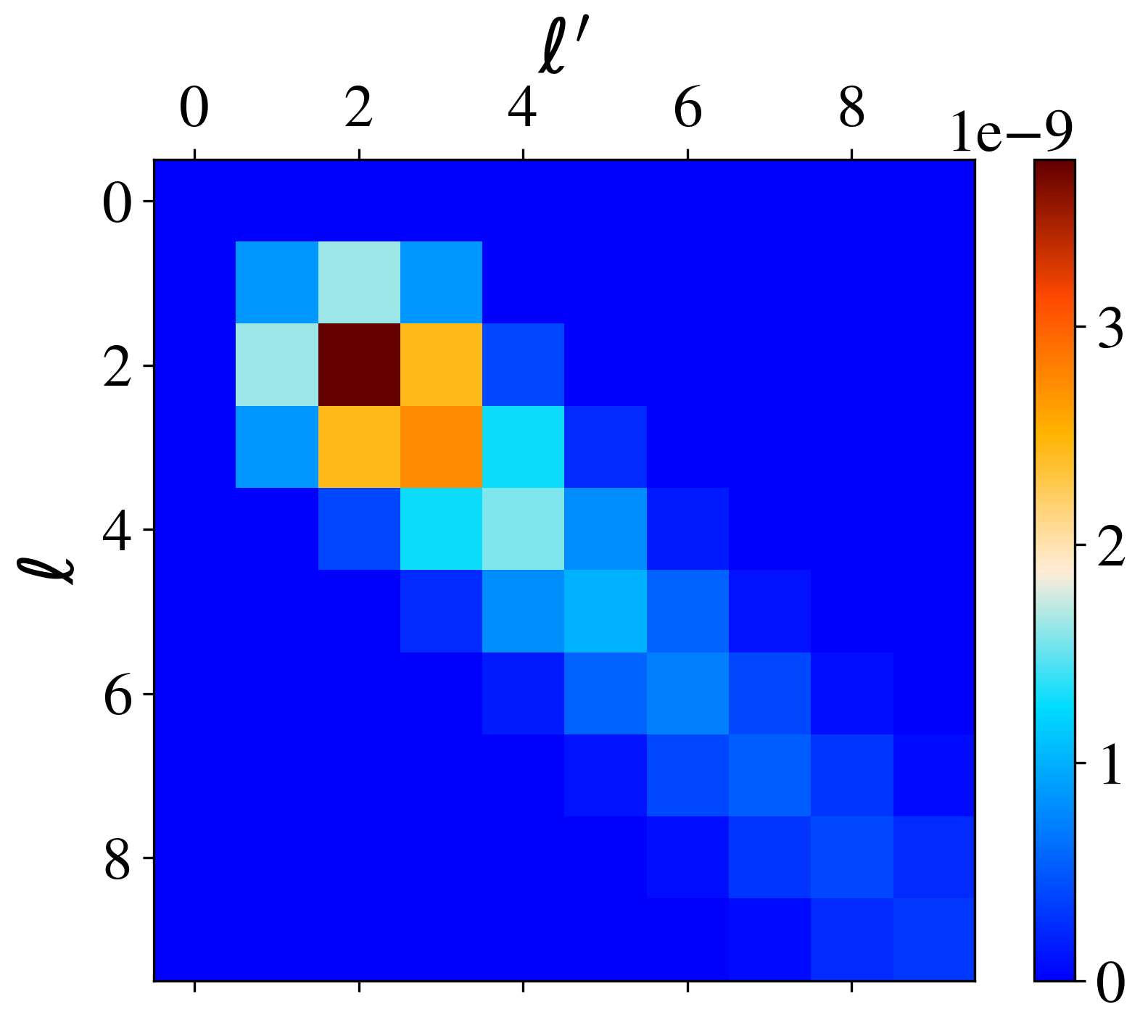}
    \caption{Covariance matrix between $\ell$ and $\ell^\prime$ for $m=0$ for the first 10 multipoles of the CGWB. Here the generated variance on the dipole is clearly visible.}
    \label{fig:corr_matrix}
\end{figure}

\begin{figure}
    \centering
    \includegraphics[width=.6\hsize]{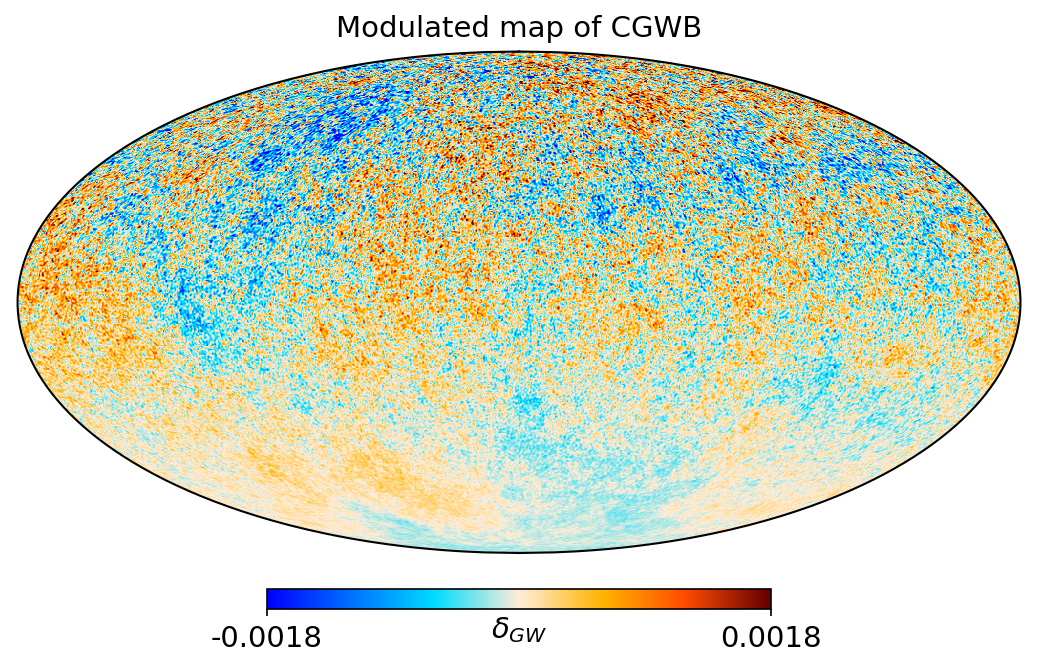}
    \caption{Modulated CGWB map with $\ell_{max} = 400$ with a very high modulation amplitude to enhance its effect.}
    \label{fig:CGWB_map}
\end{figure}

As an example, we show in figure \ref{fig:corr_matrix} the correlation matrix between $\ell$ and $\ell^\prime$ for the first 10 multiples for $m=0$ of the CGWB when we assumed $\omega_1 = 4$.
One can notice that indeed there is a small contribution to the ``pure'' dipole coming from the second order term in the modulating field and a much bigger one coming from the cross-correlation between the quadrupole and the dipole ($(\ell,\ell^\prime) = (1,2)$ or vice versa).
Setting again the modulating amplitude to $\omega_1 = 4$ just for illustrative purpose, we obtain figure \ref{fig:CGWB_map} for the modulated case of the CGWB.
Now, we compute the cross-correlation coefficient between the CGWB and CMB, defined as $\ensuremath{\mathcal C}^{CMB\times CGWB}_\ell/\sqrt{\ensuremath{\mathcal C}^{CMB}_\ell \times \ensuremath{\mathcal C}^{CGWB}_\ell}$. This coefficient will clearly depend on the value of the spectral tilt $n_t$, thus it is worth showing its variation for different values of the tilt.
Once again, we choose $n_t = \qty{-0.76,\ -0.35,\ 0,\ 0.25,\ 0.52}$.
\begin{figure}
    \centering
    \includegraphics[width=.5\hsize]{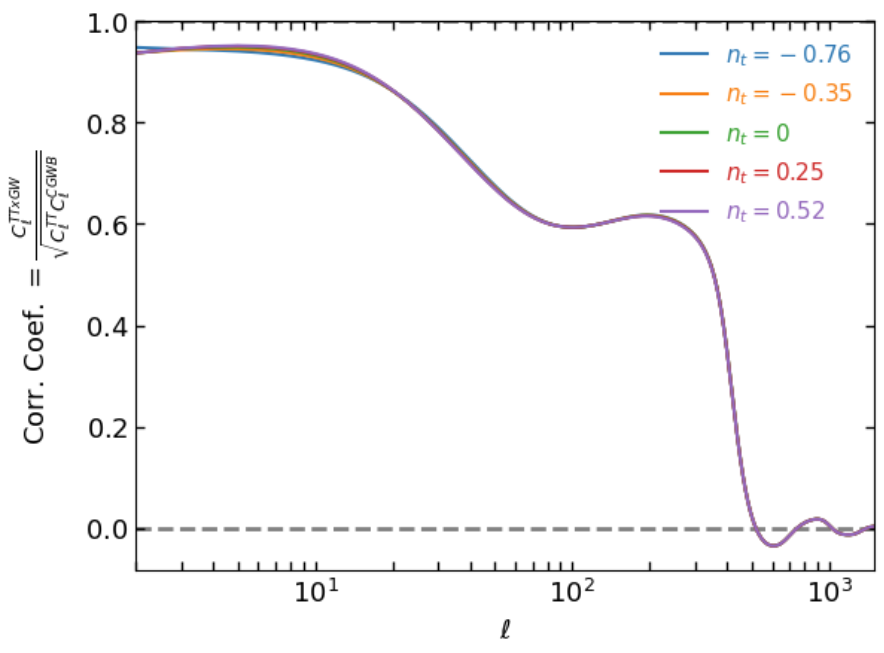}
    \caption{Correlation coefficient between the CMB and CGWB when we assume different values of the tensor spectral tilt $n_t$.}
    \label{fig:corr_coef}
\end{figure}
Figure \ref{fig:corr_coef} shows that there are two main regions where the CMB and the CGWB are correlated, namely the largest scales $\ell < 100$ and around $\ell = 200$. Indeed, the SW effect of GWs and CMB photons are generated respectively at $\eta_{in}$ and $\eta_{rec}$ (respectively through the convolution with a Dirac's delta and the visibility function). However, the spatial separation of the events is much smaller than the scale of the perturbation considered on those multipoles, thus the cross-correlation is still high\footnote{In fact, the more you go towards small scales, the more that contribution fades away, given that the spatial separation gets too large w.r.t. the scale of the perturbation.}. Instead, the other region is due to the correlation between the first acoustic peak of CMB and the ISW effect on GWs \cite{Ricciardone_2021}.

\subsection{Correlation CMB vs CGWB}
To fully exploit the cross-correlation between CMB and CGWB and the available data for the temperature field, we generate constrained realizations of the CGWB \cite{Hoffman_1991, Ricciardone_2021}. In fact, looking again at figure \ref{fig:corr_coef}, one can notice that on multipoles lower than $\approx 50$ the realization of the CGWB should be nearly deterministically fixed by our observation of CMB, because the correlation coefficient is nearly 1. 
However, in our case we have to emphasize a peculiarity: a constrained realization of the CGWB ($\Gamma_{\ell m}$) given a realization of the CMB temperature field ($a_{\ell m}$) is customarily given by \cite{chiocchetta}
\begin{equation}
    \Gamma_{\ell m} = \frac{\ensuremath{\mathcal C}_\ell^{CMB\times CGWB}}{\ensuremath{\mathcal C}_\ell^{CMB}}a_{\ell m} + \xi_{\ell m} \sqrt{ \ensuremath{\mathcal C}_\ell^{CGWB} - \frac{\qty(\ensuremath{\mathcal C}_\ell^{CMB\times CGWB})^2}{\ensuremath{\mathcal C}_\ell^{CMB}}} \ ,
\label{eq:old_contrained_real}
\end{equation}
where $\xi_{\ell m}$ is a Gaussian random field with mean 0 and unitary variance. In other words, for the multipoles where the cross-spectrum is high, the realization of the CGWB acquires a mean similar to the CMB one and will have a suppressed variance around that. Vice versa, when the cross-spectrum is low, the CGWB realizations will go back to the standard case, i.e. it will have a null mean and a variance equal to the square root of the angular power spectrum.
In spite of this, we know that the dipolar modulation model we adopted to describe the power asymmetry introduces a coupling between different multipoles, which is not accounted for in Eq.\ref{eq:old_contrained_real}. For this reason, we need to generalize this equation to a case with non-zero contributions in the off-diagonal elements of the covariance matrices. This is done by considering the known formulas of the mean and the variance of a conditioned multivariate Gaussian, which recast Eq.\ref{eq:old_contrained_real} to \cite{Marriott_1984, Bucher:2011nf}
\begin{equation}
    \Vec{\Gamma}_{m} = \ensuremath{Cov}_m^{CMB\times CGWB}\qty(\ensuremath{Cov}_m^{CMB})^{-1} \Vec{a}_{m} + \Vec{\xi}_{m}\times Chol.\qty[M_m]\ ,
\label{eq:contrained_real}
\end{equation}
\begin{equation}
    M_m =Cov_{m}^{CGWB} - Cov_{m}^{CMB\times CGWB}\qty(Cov_{m}^{CMB})^{-1}\qty(Cov_{m}^{CMB\times CGWB})^T\ .
\end{equation}
In Eq.\ref{eq:contrained_real} we are fixing the index $m$ and for each we define $\Vec{a}_{m}$, $\Vec{\xi}_m$ and $\Vec{\Gamma}_{m}$, which are vectors long $\ell_{max}-m$. Indeed, $\Gamma_{\ell m},a_{\ell m},\xi_{\ell m}$ of Eq.\ref{eq:old_contrained_real} are the elements of these vectors: $\Vec{a}_{m}$ is extracted from the complete vector of the CMB realization fixing $m$, whereas $\Vec{\xi}_m$ is a multivariate Gaussian vector with 0 mean and unitary variance ($\mathbb{I}_{\ell_{max}-m}$). Finally, $\Vec{\Gamma}_{m}$ is built exploiting the covariance matrices between $\ell, \ell^\prime$ (Eq.\ref{eq:total_correlator} shows the CGWB case, represented graphically in figure \ref{fig:corr_matrix}). Also, $Chol.$ indicates that we are taking the Cholesky decomposition of the matrices $M_m$ in order to get the correct covariance. Thus, since we want to assess the significance of a certain modulating amplitude using these constrained realizations, we will plug in the expression of the covariance matrices the estimated value of the amplitude we can get from our CMB observation (see section \ref{sec:res_asses} for further details). Furthermore, one can easily show that in the case of diagonal matrices (i.e. no modulation), Eq.\ref{eq:contrained_real} is equivalent to Eq.\ref{eq:old_contrained_real}.

\begin{figure}
    \centering
    \includegraphics[width=.49\hsize]{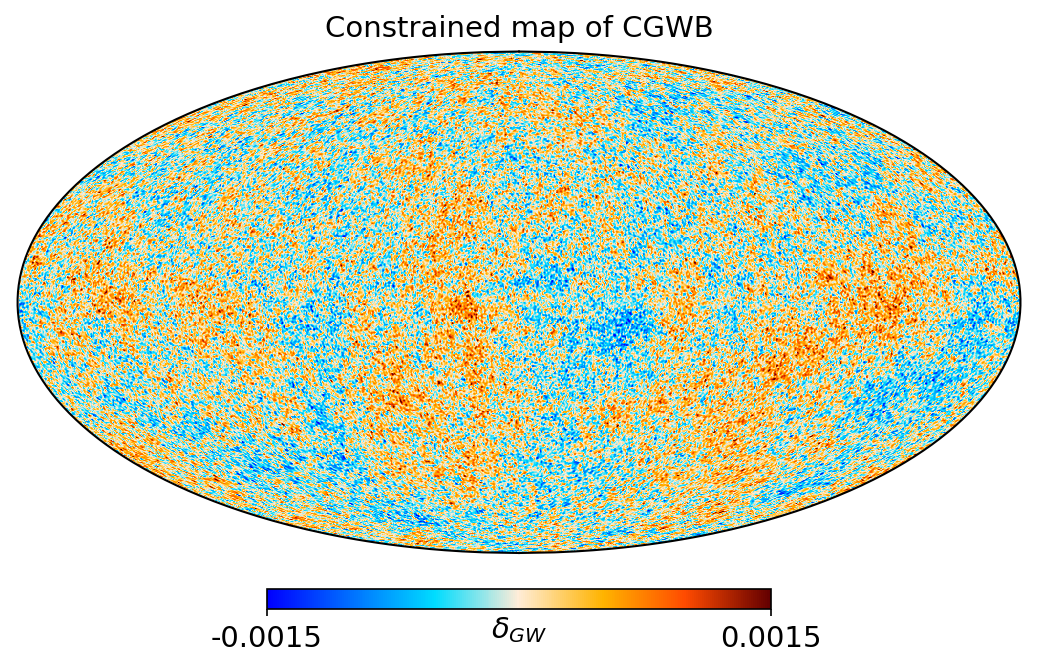}
    \hfill
    \includegraphics[width=.49\hsize]{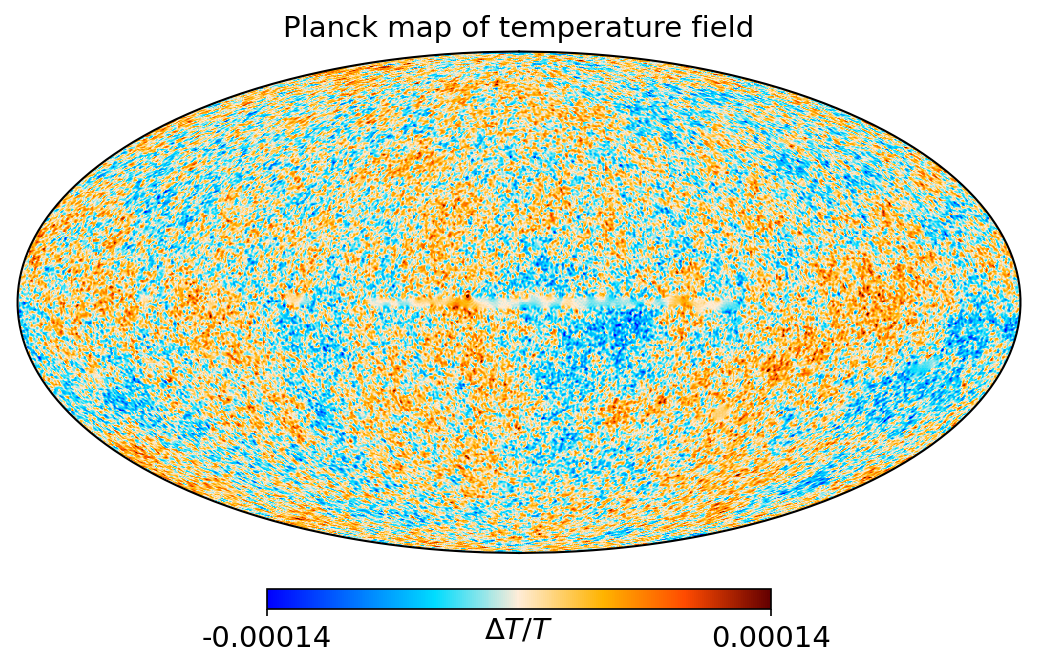}
    \caption{Constrained realization of the CGWB with $\ell_{max} = 400$ (left) and Planck CMB map, whose  spherical harmonics coefficients with $\ell > \ell_{max} = 400$ are filtered out (right). It is clear that the large-scale features of Planck's map are faithfully reproduced here, whereas the two observables drift apart on smaller scales.}
    \label{fig:cons_cgwb}
\end{figure}
Comparing the left and right panels of figure \ref{fig:cons_cgwb}, the latter being Planck's temperature map downgraded to $\ell_{max} = 400$, one can notice that the large-scale features present in the CMB are faithfully reproduced in our realization of the CGWB, given that the cross-correlation is high on those scales. Let us recall that in order to pass to a energy-contrast-description of the CGWB, we multiply Eq.\ref{eq:contrained_real} by $(4-n_t)$.
\section{The role of GWs in assessing the anomaly} \label{sec:res_asses}
\subsection{Estimator of the modulating amplitude}
We apply the estimator defined in \cite{Hu} to assess the amplitude of the modulating field from a sky-map realization. 
For every couple of sky-realizations $X_{\ell m}^{*} Y_{\ell+1, m}$, where $XY = \qty{TT,\ GWGW,\ TGW,\ GWT}$, they define the estimator for $\omega_1$ as
\begin{equation}
    \hat{\omega}_{1, \ell m}^{X Y}=\frac{X_{\ell m}^{*} Y_{\ell+1, m}}{f_{\ell}^{X Y} R_{\ell+1, m}^{1 \ell}}\ .
\label{eq:estim_1}
\end{equation}
The coefficients $f_\ell^{XY}$ are given by
\begin{equation}
    \expval{X_{\ell m}^{*} Y_{\ell+1, m}} = \omega_1 f_\ell^{XY} R_{\ell+1, m}^{1 \ell}\ , 
\end{equation}
so that $\expval{\hat{\omega}_{1, \ell m}^{X Y}} = \omega_1$. In our case the expression of these coefficients is fairly simple:
\begin{equation}
    f_\ell^{XY} = \ensuremath{\mathcal C}_\ell^{XY} + \ensuremath{\mathcal C}_{\ell+1}^{XY}\ ,
\end{equation}
with $f_\ell^{TGW} = f_\ell^{GWT}$.

The estimators are combined to form a joint estimator defined as
\begin{equation}
    \hat{\omega}_1 = \sum_{XY} \sum_{\ell m} A_{\ell m}^{X Y} \hat{\omega}_{1, \ell m}^{X Y}\ ,
\end{equation}
where $A_{\ell m}^{X Y}$ are weights which satisfy
\begin{equation}
    \sum_{XY} \sum_{\ell m} A_{\ell m}^{X Y} = 1
\end{equation}
and minimize the variance of the joint estimator. Their definition is as follows
\begin{equation}
A_{\ell m}^{X Y}=\frac{\sum_{X^{\prime} Y^{\prime}}\left[\mathcal{D}^{(\ell)}\right]_{X Y, X^{\prime} Y^{\prime}}^{-1}\left(R_{\ell+1, m}^{1 \ell}\right)^{2}}{\sum_{X Y, X^{\prime} Y^{\prime}} \sum_{\ell m}\left[\mathcal{D}^{(\ell)}\right]_{X Y, X^{\prime} Y^{\prime}}^{-1}\left(R_{\ell+1, m}^{1 \ell}\right)^{2}}\ ,
\end{equation}
where
\begin{equation}
    \mathcal{D}^{(\ell)}_{X Y, X^{\prime} Y^{\prime}} = \frac{\ensuremath{\mathcal C}_\ell^{XX^\prime}\ensuremath{\mathcal C}_{\ell +1}^{YY^\prime}}{f_\ell^{XY}f_\ell^{X^\prime Y^\prime}}\ .
\label{eq:noiseless}
\end{equation}
As a validating step of this estimator and our machinery, we generate a set of $N = 1000$ realizations of the CMB and CGWB. We modulate each CMB realization with $\omega_1 = 0.5$ and from each of them we generate a constrained realization of the CGWB. This is the perfect place to emphasize the difference between using Eq.\ref{eq:old_contrained_real} or Eq.\ref{eq:contrained_real} to produce the constrained realizations of the CGWB.
\begin{figure}
    \centering
    \includegraphics[width=.49\hsize]{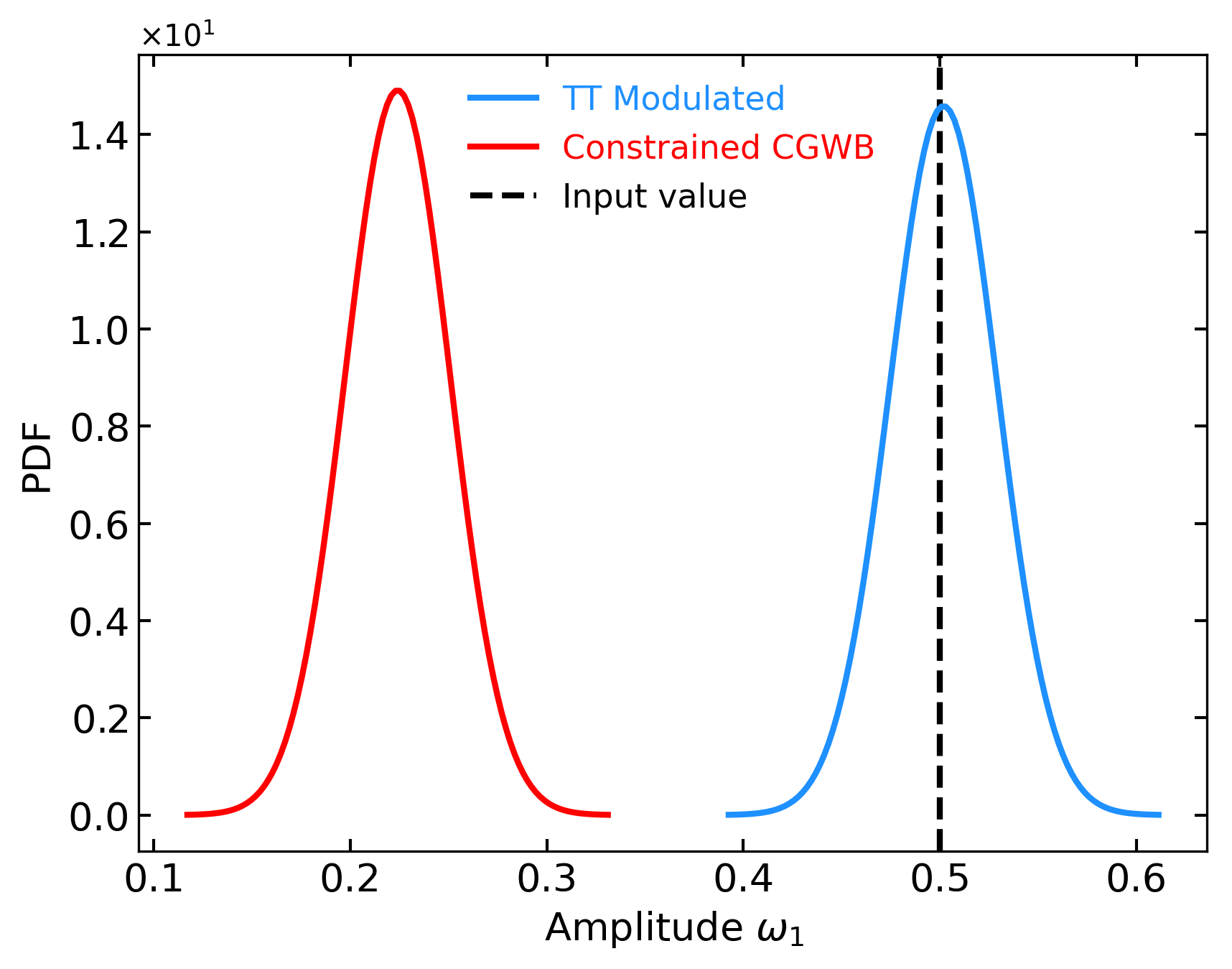}
    \hfill
    \includegraphics[width=.49\hsize]{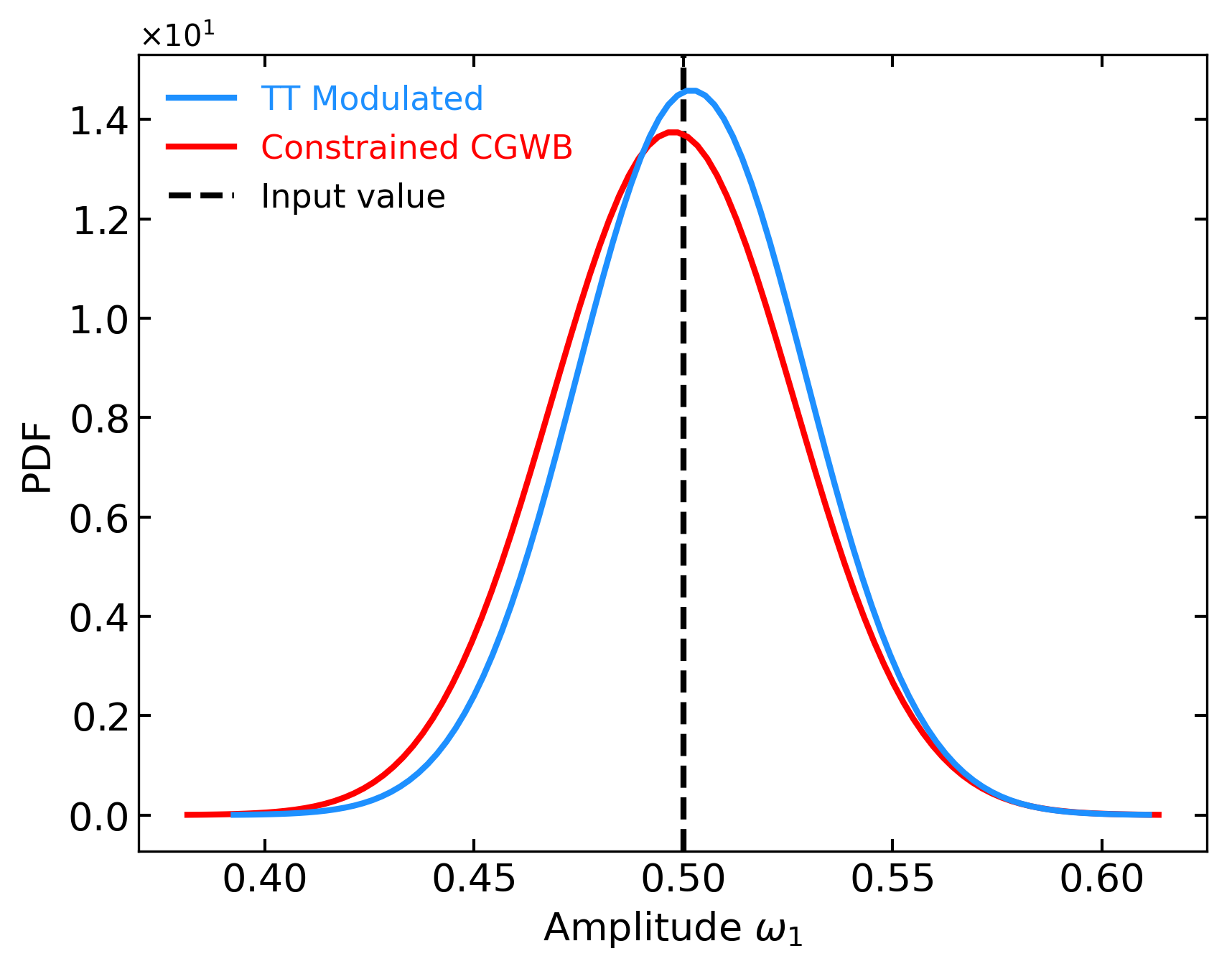}
    \caption{Estimated modulating amplitude for a set of 1000 realizations, assuming $\ell_{max} = 100,\ n_t=0$. The dashed black line indicates the input value of the modulating amplitude. The solid blue curve represents the CMB modulated realizations, whereas the solid red one the constrained CGWB realizations. On the left, we exploited Eq.\ref{eq:old_contrained_real}, whereas on the right its generalized counterpart, Eq.\ref{eq:contrained_real}.}
    \label{fig:mod_sim_estimator}
\end{figure}
The left panel of figure \ref{fig:mod_sim_estimator} shows that indeed the estimated amplitudes in the case of the CMB are correctly distributed around the input value, with some dispersion given by cosmic variance.
The distribution for the CGWB is however centered around a lower value than the CMB one. Indeed this is not surprising, because we obtain these realizations as constrained ones from the CMB set using Eq.\ref{eq:old_contrained_real}. Since the ``signal'' of the modulation is transferred to the CGWB by $a_{\ell m}$ mainly through the first $\approx 50$ multipoles, where the correlation is high, and given that the realizations we produce have $\ell_{max} = 100$, the estimation on GWs is dragged to 0. However, we underline that this is an effect solely due to our procedure. Indeed, in the right panel of figure \ref{fig:mod_sim_estimator} one can see that also the GW-estimates are distributed around our input value for the CMB modulated set with a similar variance. This is due to the fact that in this case we use Eq.\ref{eq:contrained_real}, thus even in low correlation regions, the covariance of the CGWB realization retains the coupling between multipoles induced by the presence of the modulation, which carries the modulation-amplitude-information in all multipoles.

In any case, from this point on we will set $\ell_{max} = 20$, given that the power asymmetry affects only large scales \cite{collaboration2019planck}. Also, we will be using the generalized expression of the constrained realizations, i.e. Eq.\ref{eq:contrained_real}, given that it best represents our model predictions.

\subsection{Cosmic variance distribution}
We can now study what the CV distribution looks like when we add GWs using the joint estimator defined from Eq.\ref{eq:estim_1} to Eq.\ref{eq:noiseless}. This will allow us to find our null-hypothesis in order to assess the significance of CMB power asymmetry when adding a GWs observation. In this case we must use unconstrained realizations of the CGWB since we want both fields to fluctuate freely \cite{chiocchetta}. The expression for unconstrained realizations can be obtained from Eq.\ref{eq:contrained_real} setting $\ensuremath{Cov}_m^{CMB\times CGWB} = 0$ \footnote{So equivalently, one can obtain the correct expression from Eq.\ref{eq:old_contrained_real} while imposing $\ensuremath{\mathcal C}_\ell^{CMB\times CGWB} = 0$.}. For this analysis we assume a scale-invariant power spectrum, thus $n_t=0$. However, a clarification is due: given that we have to study how CV can simulate the presence of a modulation, we are assuming a cosmological model without any modulating field. For this reason, the $\ensuremath{\mathcal C}_\ell^{XX^\prime}$ we used for the estimator are the standard ones (e.g. Eq.\ref{eq:zeroth} multiplied by the proper $\Gamma\to\delta_{GW}$ conversion factor).
\begin{figure}
    \centering
    \includegraphics[width=.5\hsize]{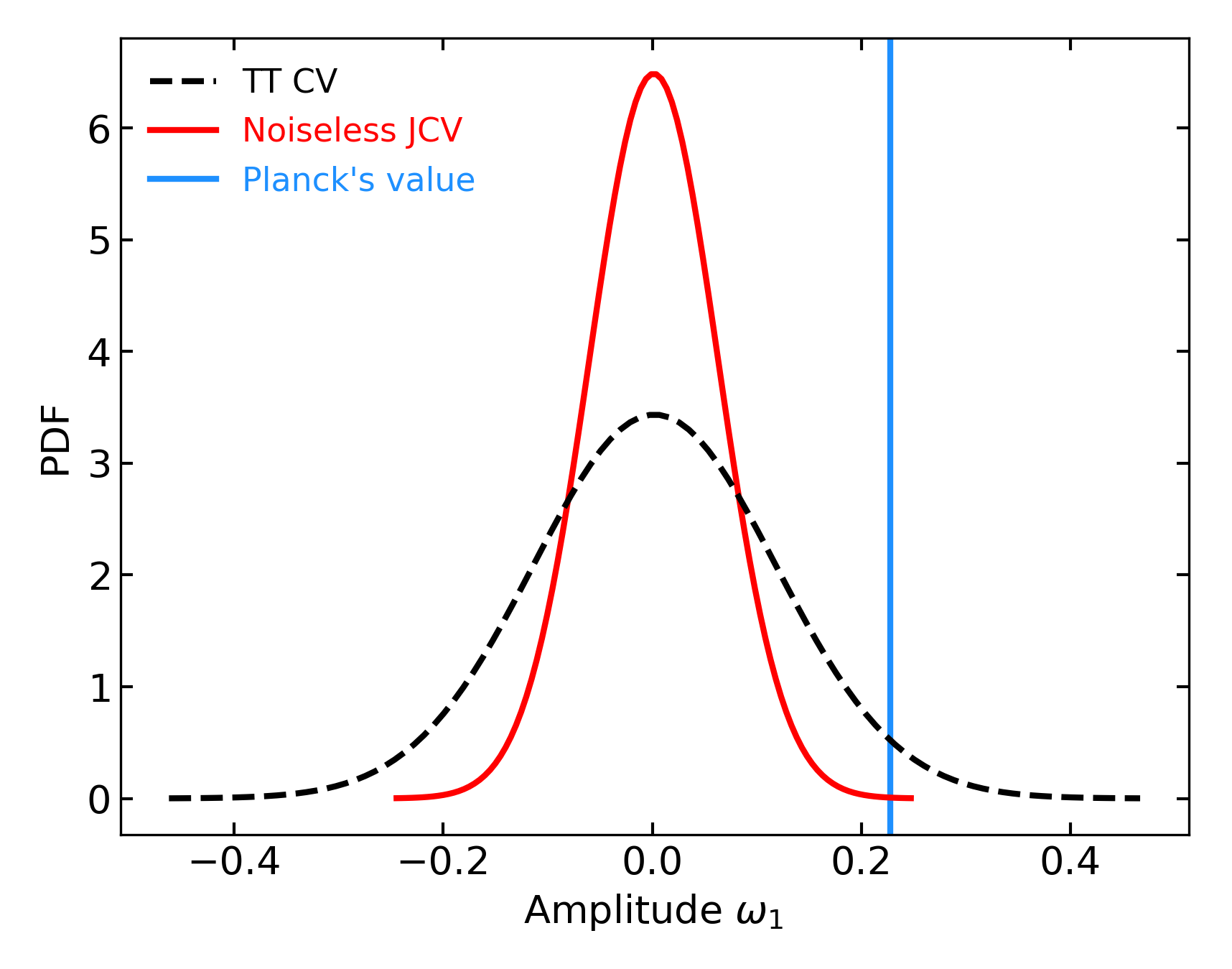}
    \caption{Estimated dispersion due to cosmic variance with a set of 1000 realizations, assuming $\ell_{max} = 20,\ n_t=0$. The dashed black curve is the dispersion for the temperature realizations, whereas the solid red line indicates the dispersion of the joint estimation using both GWs and CMB. This time the GWs realizations are obtained independently from the CMB ones. The solid blue line is the value inferred by Planck's SMICA map.}
    \label{fig:uncon_cv}
\end{figure}
Figure \ref{fig:uncon_cv} shows that the cosmic variance dispersion decreases noticeably compared to the TT CV case. From this point on, we will call JCV the shrunken distribution obtained when the CGWB is considered together with CMB temperature. The value we estimate from Planck's SMICA map $\omega_1 = 0.23$ passes from $1.95\sigma$ in the TT-only case to $3.70\sigma$ when we consider the JCV, prominently favoring this modulation model w.r.t. some unknown systematic effect. 

However, notice that we have yet to introduce any instrumental noise for the GWs part, which is expected to have an impact on the result.
Indeed, when we consider this extra source of noise Eq.\ref{eq:noiseless} is recast to
\begin{equation}
    \mathcal{D}^{(\ell)}_{X Y, X^{\prime} Y^{\prime}} = \frac{\qty(\ensuremath{\mathcal C}_\ell^{XX^\prime} + N_\ell^{XX^\prime})\qty(\ensuremath{\mathcal C}_{\ell +1}^{YY^\prime} + N_{\ell +1}^{YY^\prime})}{f_\ell^{XY}f_\ell^{X^\prime Y^\prime}}\ 
\label{eq:noisy}
\end{equation}
and we add a realization of the noise to the realization of the CGWB.

The inverse of Eq.\ref{eq:noisy}, which is the one entering the equations for the estimator, will tend to zero for noise-dominated cases. In other words, when the noise of an experiment is too large, the JCV will be completely disrupted and will become much broader than the TT CV, since the GWs part will contribute to enhance largely the dispersion of the simulations.
Here, we will consider LISA and BBO as our testing frameworks, whose instrumental noise is obtained through the publicly available code schNell \cite{schnell, Ricciardone_2021}. 
The noise spectrum of these instruments will depend on the value we assume for $n_t$. In fact, the expected noise levels $N_\ell$ of LISA and BBO are related to the angular noise spectrum of density contrast through $N_\ell^{CGWB} = N_\ell/\ols{\Omega}_{GW}^2$. In our case, having assumed a power-law description of the primordial power spectrum, we compute the average energy density at some frequency\footnote{The reference frequency of LISA is $10^{-2}$ Hz, instead for BBO $1$ Hz.} as \cite{Planck_2018}:
\begin{equation}
    \ols{\Omega}_{GW}(f) = \frac{rA_s}{24z_{eq}}\qty(\frac{f}{f_{pivot}})^{n_t}\ ,
\end{equation}
where $r$ is the tensor-to-scalar ratio at $f_{pivot} \approx 1.55\times10^{-17}$ Hz, corresponding to $k = 0.01$ Mpc$^{-1}$, and $z_{eq} \approx 3400$ is the redshift of matter-radiation equality (calculated by the Boltzmann code). In this way, the more blue tilted the spectral index is, the higher will be $\ols{\Omega}_{GW}$ at the reference frequencies of both BBO and LISA.

The resolution of GW interferometers is typically quite low and limited to the first 10-20 multipoles \cite{Bartolo:2022pez}, so the JCV we obtain will improve the TT CV in just few cases. We report here the case of BBO when we assume $n_t=0.52$. Indeed, even if choosing a higher tilt decreases the GW signal (see figure \ref{fig:CGWB_nt}), its effect on how much power reaches the typical scales of interferometers is by far more dominant. Thus, we choose to maximize the value in the allowed range of \cite{Planck_2018}.
\begin{figure}
    \centering
    \includegraphics[width=.5\hsize]{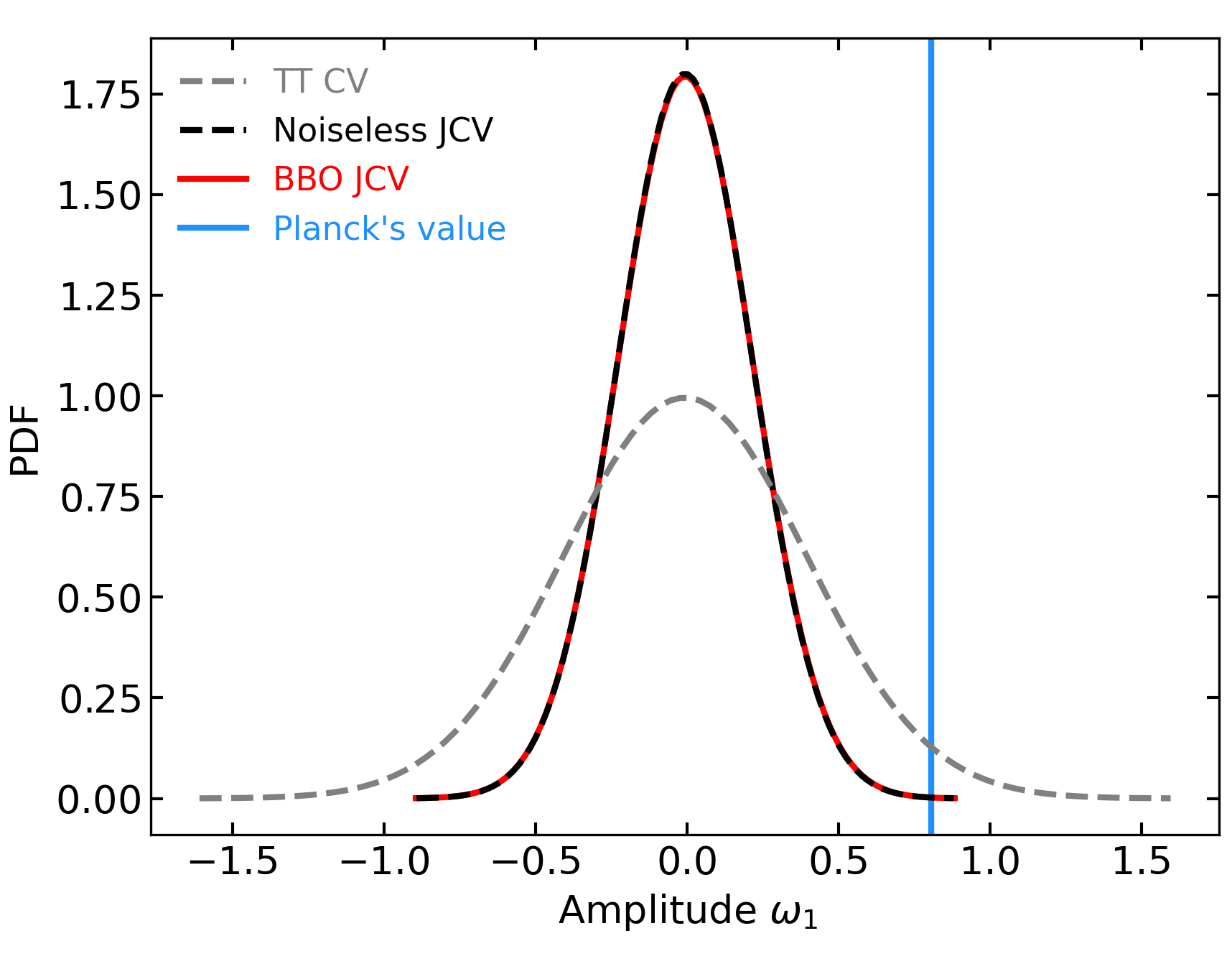}
    \caption{JCV when we include the instrumental noise coming from BBO. The red curve is obtained assuming $\ell_{max}=6,\ n_t=0.52$, whereas the blue vertical line indicate the value estimated from Planck's temperature map. The dashed gray and black lines represent respectively the CV when using only CMB temperature realizations and the noiseless JCV, thus when we include also the CGWB.}
    \label{fig:CV_nt_3}
\end{figure}
To obtain figure \ref{fig:CV_nt_3}, we limit our analysis to the first six multipoles, given that the noise for $\ell > 6$ is too large. The figure shows that in spite of the noise the GW signal is strong enough to allow BBO to be signal-dominated and to recover identically the noiseless JCV. Notice that the Planck estimated value we report here is different w.r.t. the one showed in figure \ref{fig:uncon_cv} because we changed the range of multipoles. Here the value of the modulating amplitude obtained for Planck's SMICA map is $\omega_1 = 0.80$.

\subsection{Forecasts for future observations}
Finally, we can forecast what we can expect to see with a future observation of GWs. 
Once again we start with the noiseless case and we generate a set of 1000 realizations of the CGWB constrained on the Planck's SMICA map, assuming that a modulation is already present in this map. Also here, a clarification is due: given the assumption we make for Planck's SMICA map, the modulating-information is passed to the GW background through the fact that the CGWB realizations are constrained on the CMB one. Thus, we are assuming that the model describing the anisotropies of both CMB and CGWB is a modulated one. So, the $\ensuremath{\mathcal C}_\ell^{XX^\prime}$ we use in the estimator definitions are the $\ensuremath{\mathcal C}_\ell^{(1)}$ mentioned in section \ref{sec:modulation} (e.g. see Eq.\ref{eq:first_Cell} for the CGWB equation). This is indeed a difference w.r.t. \cite{Hu}, where $\ensuremath{\mathcal C}_\ell^{(0)} = \ensuremath{\mathcal C}_\ell^{(1)} = \ensuremath{\mathcal C}_\ell^{(2)}$. In spite of this, having assumed $\ensuremath{\mathcal C}_\ell^{(1)}$ is crucial to fully capture the modulation amplitude, given that we do not neglect the ISW effect, of which we find a modified expression w.r.t. the standard one in Eq.\ref{eq:transfer_3}, and we do not assume any approximation for the gravitational potentials transfer functions.

\begin{figure}
    \centering
    \includegraphics[width=.5\hsize]{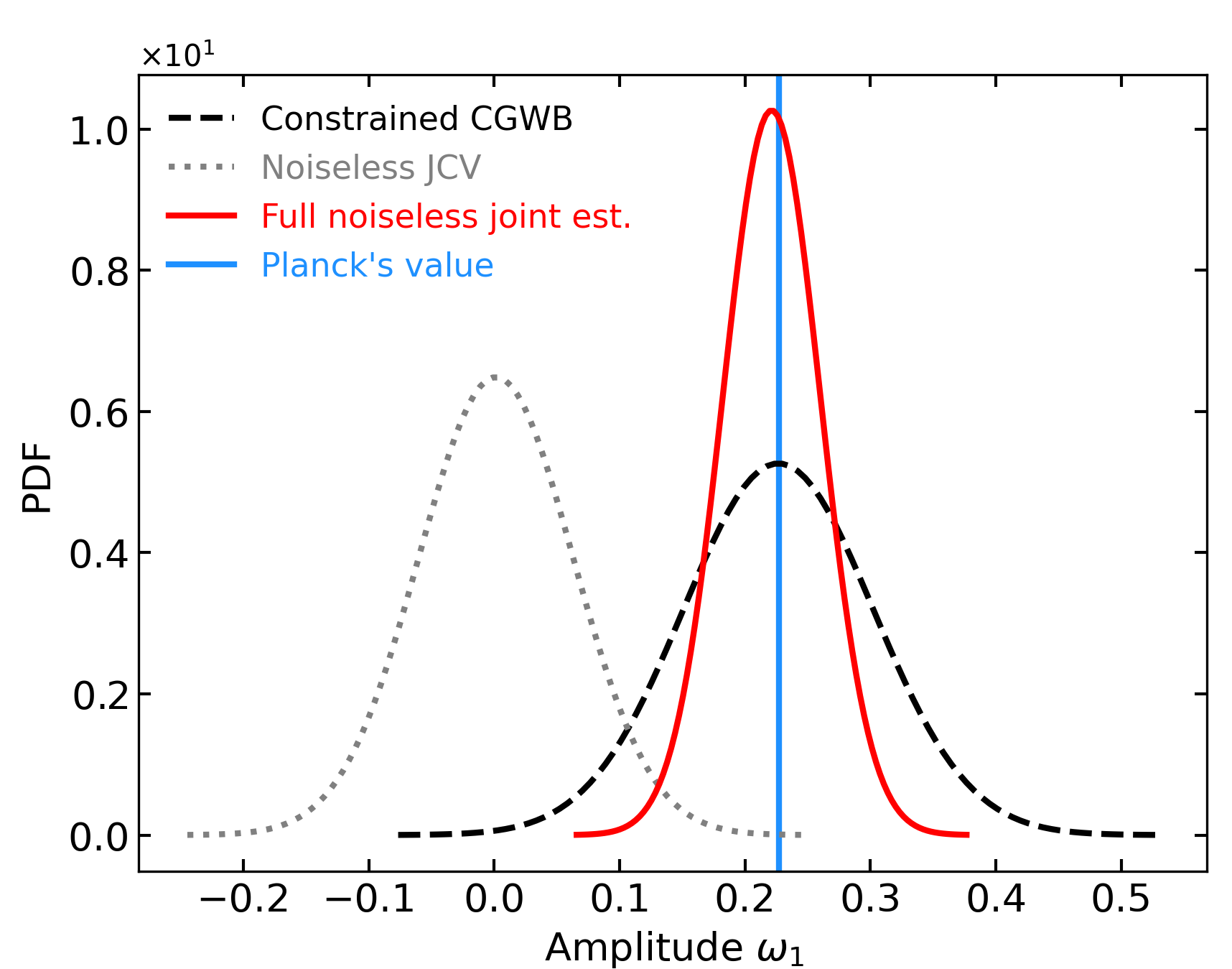}
    \caption{Estimated modulating amplitude for a set of 1000 realizations constrained on the Planck's map, assuming $\ell_{max} = 20,\ n_t=0$. The dashed black curve is the dispersion for the CGWB realizations constrained to the Planck's map, whose value is represented as a solid vertical blue line. The solid red curve indicates the dispersion of the joint estimation using both the realizations of the CGWB and Planck's map. Finally, the dotted gray line represents the JCV dispersion.}
    \label{fig:planck_estimator}
\end{figure}
Figure \ref{fig:planck_estimator} shows that if a modulation is indeed there, the CGWB will distribute as a normal with $\qty{\mu,\sigma} = \qty{0.23, 0.08}$, whereas the joint estimator with $\qty{\mu,\sigma} = \qty{0.22, 0.04}$.

\begin{figure}
    \centering
    \includegraphics[width=.5\hsize]{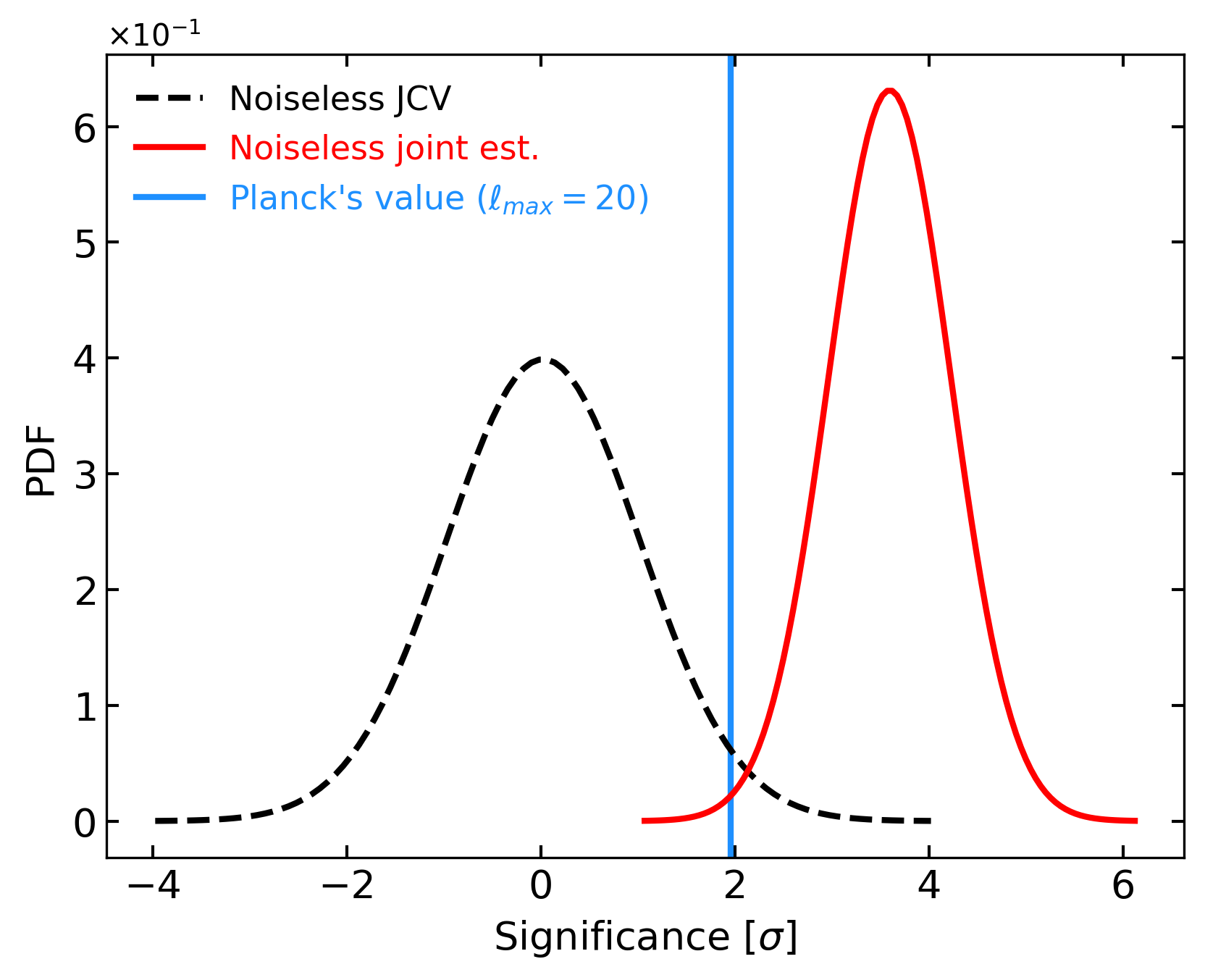}
    \caption{Estimated significance for the joint estimator of the modulating amplitude using a set of 1000 CGWB realizations constrained on the Planck's map, assuming $\ell_{max} = 20,\ n_t=0$. The dashed black curve is the significance of the JCV (red curve of figure \ref{fig:uncon_cv}). The Planck's map significance without GWs is represented as a solid vertical blue line, whereas the solid red curve indicates the dispersion of the joint estimation.}
    \label{fig:compatibility}
\end{figure}

For the sake of clarity, we compute the significance of the joint estimator w.r.t. the JCV. Figure \ref{fig:compatibility} clearly shows the increment in significance that we can expect from a signal-dominated detection of the CGWB. The joint estimator significance is distributed as a normal distribution with $\qty{\mu,\sigma} = \qty{3.60, 0.63}$; specifically, the $83.4\%$ of the simulations shown in figure \ref{fig:compatibility} below the red curve has a significance $\geq 3\sigma$ and all of them improve the significance w.r.t. the Planck's value from SMICA map when assuming $\ell_{max}=20$, i.e. $1.95\sigma$ (blue solid line).

\begin{figure}
    \centering
    \includegraphics[width=.5\hsize]{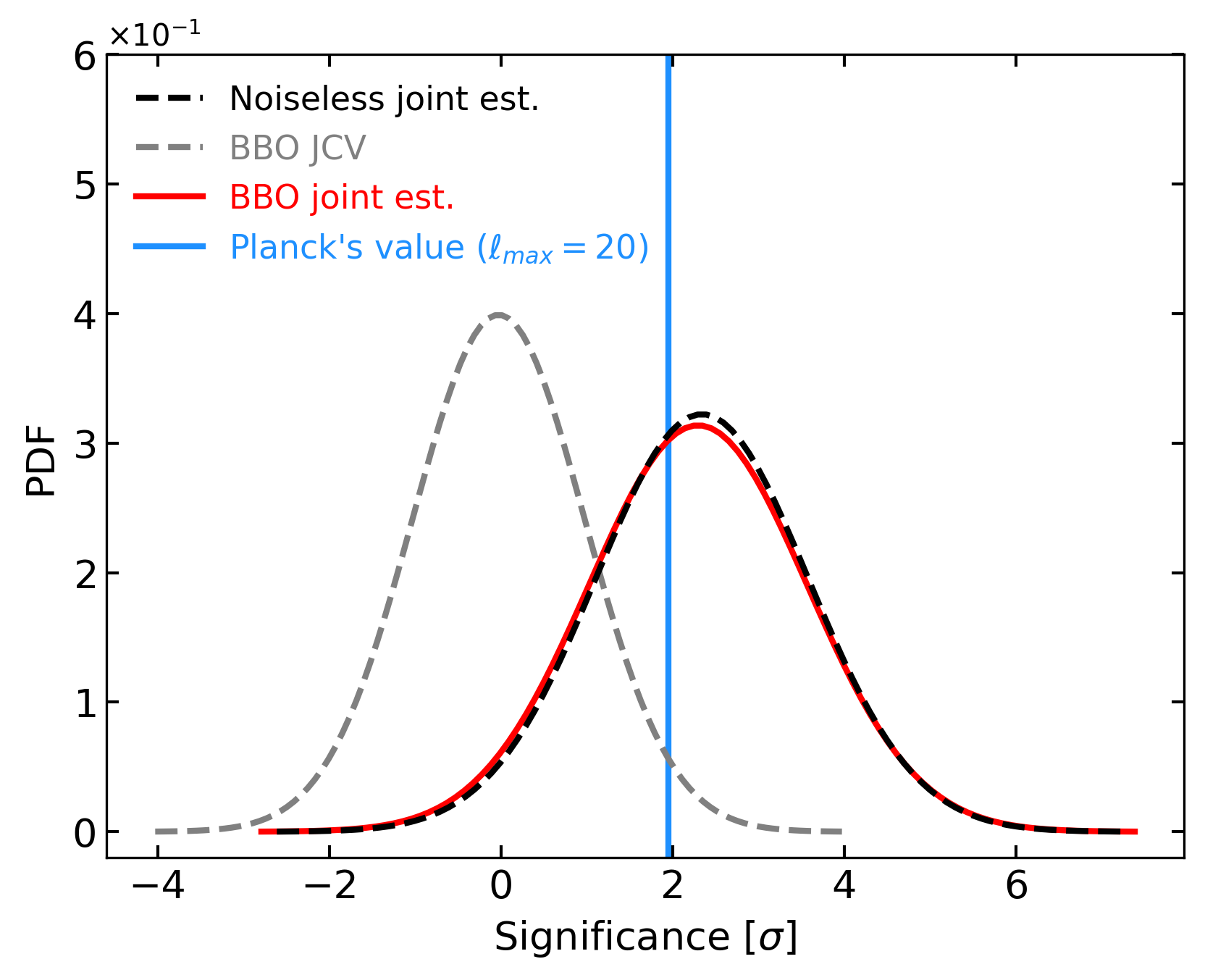}
    \caption{Estimated significance for the joint estimator of the modulating amplitude using a set of 1000 CGWB realizations constrained on the Planck's map and including the instrumental noise of BBO, assuming $\ell_{max} = 6,\ n_t=0.52$ (solid red line). The dashed gray curve is the dispersion of the BBO JCV we show in figure \ref{fig:CV_nt_3}. The Planck's map significance without GWs is represented as a solid vertical blue line. The black dashed line is the noiseless case.}
    \label{fig:BBO_forecast}
\end{figure}

At this point, we repeat the same analysis including the instrumental noise from either LISA or BBO.
As we previously mentioned, the CGWB anisotropies are very hard to observe, thus we report here only the results for BBO when assuming $n_t = 0.52$. As it turns out, all the other cases present a large noise, which quickly dominates the measurement for multipoles $\ell > 4$.
\begin{figure}
    \centering
    \includegraphics[width=.5\hsize]{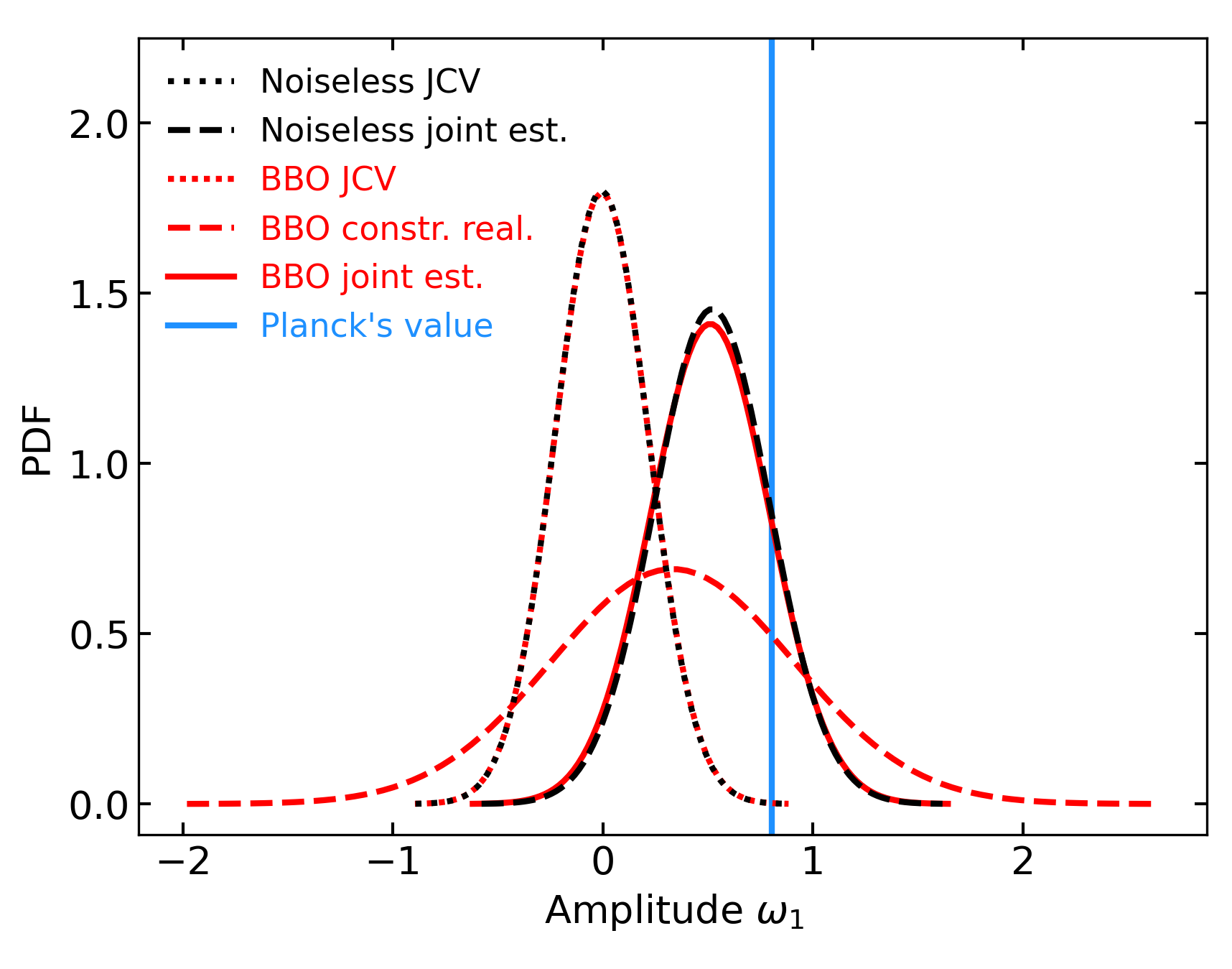}
    \caption{Joint estimation of the modulating amplitude when we include the instrumental noise of BBO, assuming $\ell_{max} = 6,\ n_t=0.52$. The dotted black and red lines are respectively obtained including or not the BBO instrumental noise. The dashed black line indicate the noiseless joint estimation from the constrained realizations on the Planck's SMICA map, whereas the dashed red one represent the CGWB constrained realizations. Finally, the red solid line is the joint estimation when we include the BBO instrumental noise.}
    \label{fig:BBO_estim}
\end{figure}
Figure \ref{fig:BBO_estim} shows the results: once again we limit our analysis to the first six multipoles. This allows BBO to recover very well the modulation inherited by the Planck's map  with the same precision of the noiseless case (compare the solid red curve with the dashed black one). As done before, we recast the BBO JCV and joint estimation in terms of significance in figure \ref{fig:BBO_forecast}. 
This shows once again the increment in significance that we can expect from a detection of the CGWB. The BBO joint estimator significance is distributed as a normal distribution with $\qty{\mu,\sigma} = \qty{2.3, 1.3}$; this time, the $27.9\%$ of the simulations shown in figure \ref{fig:BBO_forecast} below the red curve has a significance $\geq 3\sigma$. This degradation is mainly due to the fact that we had to limit $\ell_{max} = 6$. Indeed, in the noiseless case (black dashed line) about the same percentage (within $1\%$) of the simulations has a significance $\geq 3\sigma$ . Also, the $60.1\%$ of the BBO simulations improves the significance w.r.t. to the Planck's value of $1.95 \sigma$.

So in spite of this degradation, it is remarkable that BBO has indeed the potential to shed light on the significance of the CMB power asymmetry.

\section{Conclusions}\label{sec:conclusions}
CMB anomalies have been observed since WMAP and have been reassessed by the Planck satellite. Specifically, the CMB power asymmetry, i.e. a difference in the mean power over the two different hemispheres centered around $(l,b) = (221, -20)$, has an estimated significance of $1.95\sigma$. This suggests a possible breaking of statistical isotropy on large-scales, which has precise and testable consequences on CMB for each physical model trying to describe such an anomaly. These effects reflect also on the CGWB, a well-known prediction of inflationary models, through the same Boltzmann equations regulating both the propagation of gravitons and photons.

We have computed the effect on the two-point correlation functions of both the CMB temperature field and the CGWB in the case of a dipolar modulation of the gravitational potentials, which can potentially reproduce this power asymmetry. This modulation breaks statistical isotropy on our Hubble volume without flawing the Universe's global isotropy and homogeneity.

These kinds of models generate a coupling between multipoles $\ell$, $\ell \pm 1$ and $\ell \pm 2$ in all observables, which can then be tested through the statistical tools developed in \cite{Hu}. GWs behaves nearly identically to CMB temperature, thus we exploited constrained realizations of the CGWB to perform our joint analysis with the SMICA Planck temperature map and unconstrained realizations of both the CMB and the CGWB to assess the cosmic variance associated to a model with no modulation. 

In section \ref{sec:modulation}, we present the main theoretical results of this work: the analytic expression of CMB and CGWB anisotropies in the presence of a dipolar modulation of the gravitational potentials. In particular, we included the SW and ISW for both observables. In section \ref{sec:res_plot}, we explore the dependency of the CGWB angular power spectrum on the tensor spectral tilt $n_t$ (also, in appendix \ref{app:SvsT} we explore this dependency distinguishing between scalar and tensor contributions to the anisotropies). This is done in a similar fashion as in
\cite{valbusa_2020} for other parameters as $N_{eff}$ (see also
\cite{Braglia_2021}).
In section \ref{sec:res_asses}, we study the role of GWs in assessing the significance of the CMB power asymmetry. We show that in the noiseless case, the significance gets severely increased as shown in figure \ref{fig:compatibility}. Specifically, the $83.4\%$ of the simulations has a significance $\geq 3\sigma$ and all of them improve the value from Planck's map of $1.95\sigma$. Also, we study the capabilities of both LISA and BBO to observe this kind of anomaly. 
In the case of a blue tilted spectrum ($r = 0.066$ at $0.01$ Mpc$^{-1}$ and $n_t = 0.52$, saturating the upper bounds of \cite{Planck_2018}), useful to provide enough energy density of GWs at the scales of the considered interferometers, we show that BBO has the ability to fully reproduce the noiseless case, given that it is signal-dominated. Indeed, when limiting to the first six multipoles, the $60.1\%$ of the simulations improves the significance we obtain for the Planck's temperature map, i.e. $1.95\sigma$, and the $27.9\%$ reaches a significance greater than $3\sigma$. This suggests that a future observation of the CGWB could be the keystone to finally assess the physical origin of the CMB power asymmetry.

It should be emphasized that we performed our analysis assuming a standard power-law characterization of the tensor power spectrum. This conservative perspective limits quite drastically the range of possibilities we could explore, especially for what regards the values of $n_t$ and so the average energy density of GWs we obtained at the typical frequencies of LISA and BBO. 
In a more general context one could study other inflationary models bringing much more power to interferometric scales and so enhancing the possibility to measure the anisotropies of the CGWB. 
Furthermore, there are a plethora of other cosmological phenomena that could produce a CGWB that we did not even mention, such as phase transitions, cosmic strings or reheating. All of these would contribute to a set of new features, which could potentially enhance the detection of GWs anisotropies.
Last but not least, in principle also the AGWB is affected by the presence of a modulation of the gravitational potentials opening another window on the exploration of this anomaly. On the same line of reasoning, we could also investigate possible connections with large-scale structures and tracers \cite{Secrest_2021,Colgain:2022xcz}. We leave this for future exploration.

\begin{appendix} 
\section{Scalar vs tensor contributions to the anisotropies} \label{app:SvsT}
As mentioned in section \ref{sec:boltzmann}, part of both the CMB and the CGWB anisotropies are induced by the presence of tensor perturbations of the metric. Here we explore the role of both scalar and tensor contributions to the anisotropies of the CGWB, the CMB and their cross-correlation, emphasizing their dependency on the tensor spectral tilt $n_t$.
\begin{figure}
    \centering
    \includegraphics[width=.49\hsize]{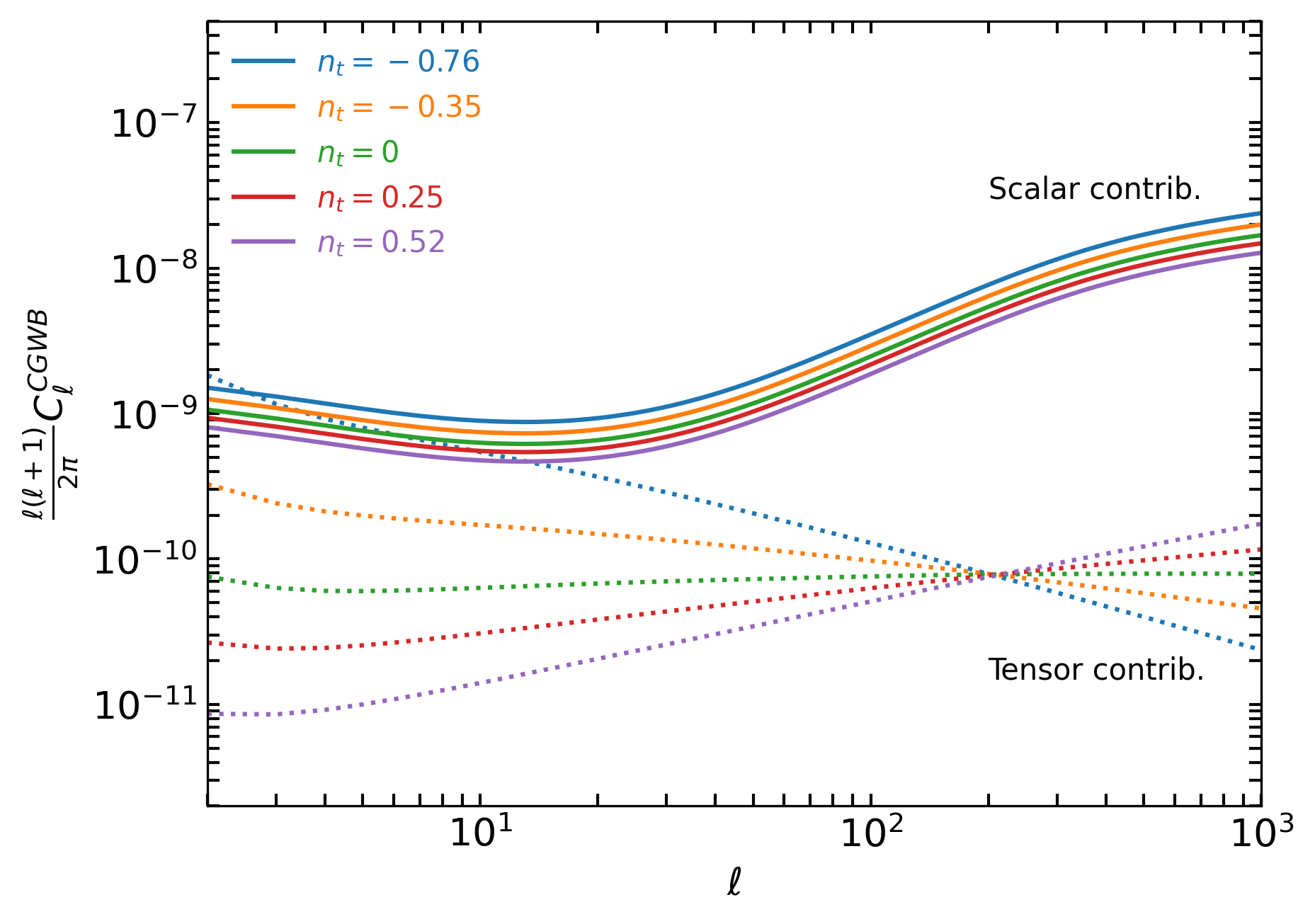}
    \hfill
    \includegraphics[width=.49\hsize]{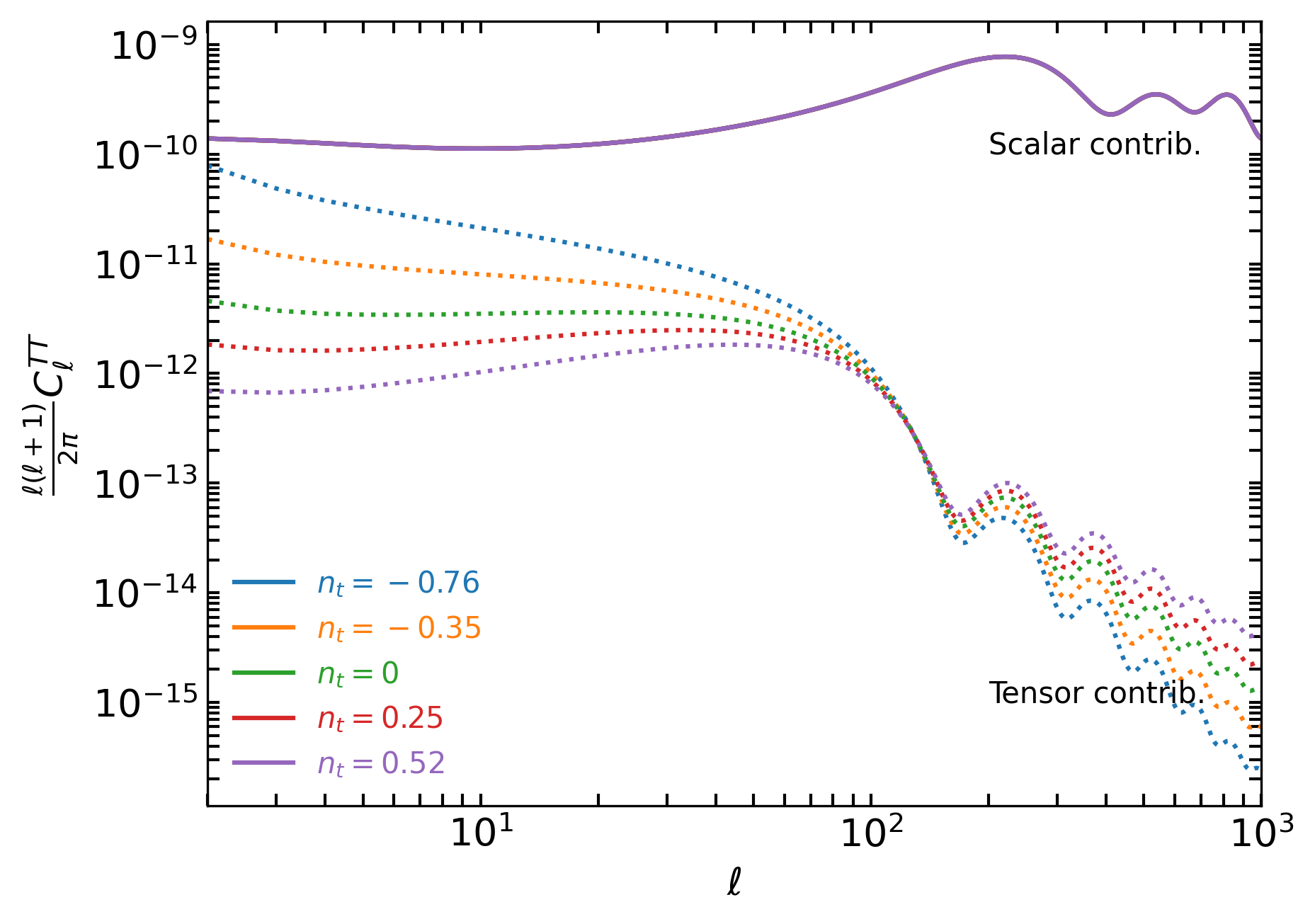}
    \caption{CGWB (left) and CMB temperature (right) angular power spectrum when we assume different values for the tensor spectral tilt $n_t$ and we do not consider any effect due to the modulation ($\omega_1=0$). Scalar induced anisotropies are reported in solid lines, whereas the tensor induced ones with dotted lines.}
    \label{fig:CGWB_ST}
\end{figure}
\begin{figure}
    \centering
    \includegraphics[width=.49\hsize]{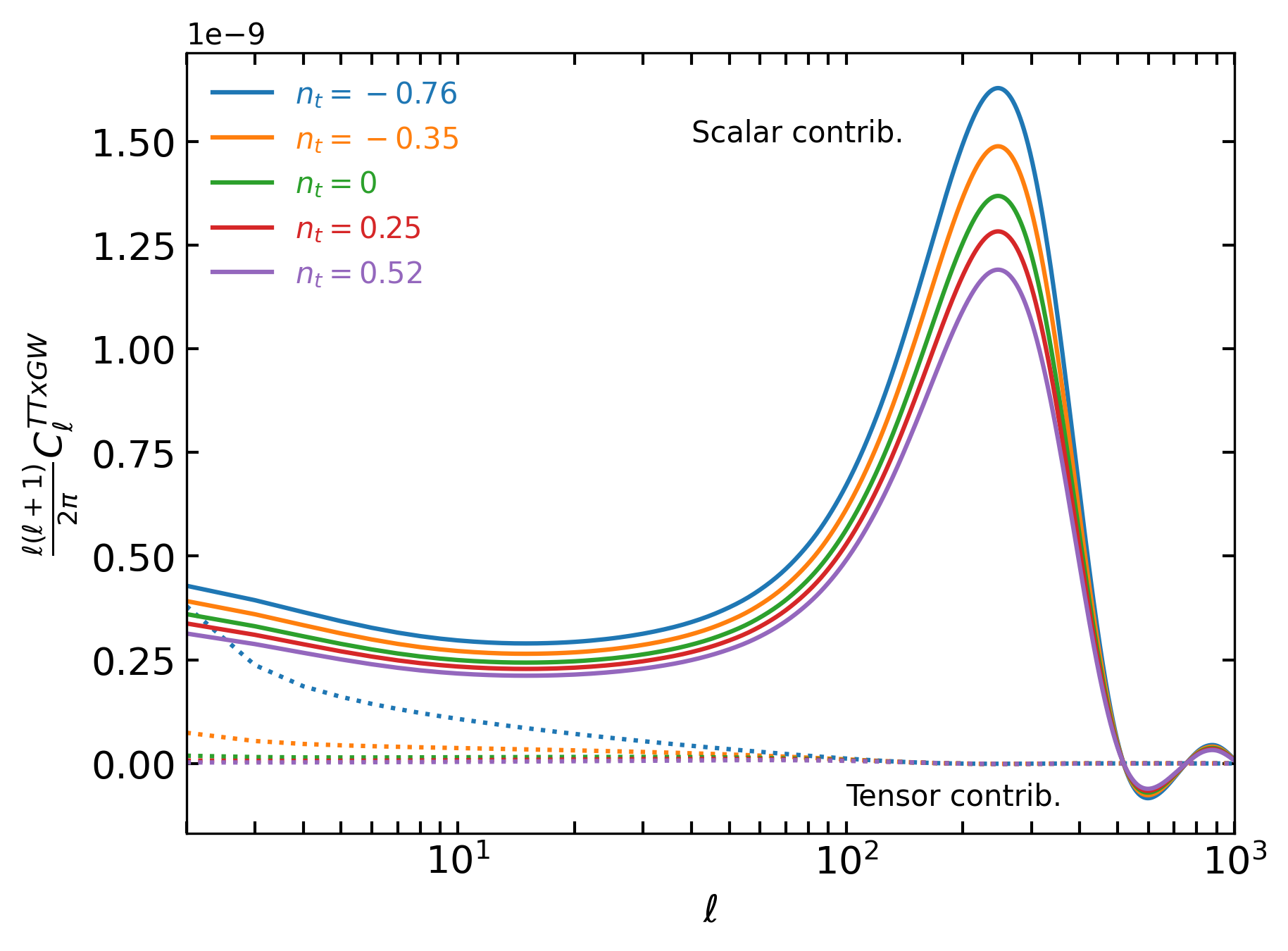}
    \caption{Cross-correlation CMBxCGWB angular power spectrum when we assume different values for the tensor spectral tilt $n_t$ and we do not consider any effect due to the modulation ($\omega_1=0$). Scalar induced anisotropies are reported in solid lines, whereas the tensor induced ones with dotted lines.}
    \label{fig:TTxGW_ST}
\end{figure}
In the case of CMB, the solution of the Boltzmann equation for the tensor modes will be proportional to the integral of a source function defined as \cite{Seljak_1996, Shiraishi_2013}
\begin{equation}
    S^{(T)}_{Temp.} = -\chi^\prime e^{-\kappa} + pol.\ ,
    \label{eq:source_cmb}
\end{equation}
where $(T)$ stands for tensor, $\kappa$ is the optical depth and $pol.$ stands for the contribution of polarization on which we are not interested (see Sec.3.6 of \cite{Shiraishi_2013} for further detail). On the other hand, the analogous solution in the case of the CGWB is proportional to \cite{Bartolo_2019}
\begin{equation}
    S^{(T)}_{GW} = -\chi^\prime \ .
    \label{eq:source_cgwb}
\end{equation}
These source functions will then play a role in the solutions for either $\Theta_{\ell m}$ or $\Gamma_{\ell m}$. However, let us recall that in order to characterize the statistical behavior of $\Theta,\Gamma$ we turn to their angular power spectrum, given that they are null-mean fields. The expression for their spectra will then contain the square of the appropriate source function, since $\expval{X_{\ell m} X^*_{\ell^\prime m^\prime}} \propto C_{\ell}^{XX}$ with $X = {\Theta,\Gamma}$. Then, Eq.\ref{eq:source_cmb} and Eq.\ref{eq:source_cgwb} generate a positive contribution of the angular power spectrum $\propto \qty(h^\prime)^2$. 

Furthermore, the $C_{\ell}^{XX}$ will also depend on the primordial tensor power spectrum, which we parametrize as a power-law (see Se.\ref{sec:res_plot}). Indeed, Fig\ref{fig:CGWB_ST} shows that a negative spectral index $n_t$ (red tilt) enhance the tensor anisotropies on large scales, which actually reach the same order of the scalar induced anisotropies. From the right panel of figure \ref{fig:CGWB_ST}, it is clear that in the case of the CMB only the tensor contribution will depend on $n_t$ through the primordial spectrum, as expected. Instead for the CGWB, the left panel of figure \ref{fig:CGWB_ST} allows us to distinguish the dependence of the scalar (see section \ref{sec:res_plot} for more details) and the tensor contributions.

The same reasoning holds also for the cross-correlation between CMB and CGWB. Indeed, its angular power spectrum contains the product of the source functions coming from CMB and CGWB. Once again the result is a positive contribution ($\propto \qty(\chi^\prime)^2$) and the overall spectrum depends on the primordial power spectrum of tensor perturbations, thus on the spectral index $n_t$. Figure \ref{fig:TTxGW_ST} shows this dependence for both the scalar and tensor contributions.

Finally, all of these spectra and their dependencies on $n_t$ generate what we see in figure \ref{fig:corr_coef}, where the tensor induced anisotropies are contributing to increase the correlation coefficient. 

\end{appendix}

\acknowledgments

We thank Alessandro Gruppuso, Eoin \'O. Colg\'ain, Frode Kristian Hansen, Lorenzo Valbusa Dall'Armi, Paolo Natoli and Roya Mohayaee for useful comments and discussions. This work is based on observations obtained with Planck (http://www.esa.int/Planck), an ESA science mission with instruments and contributions directly funded by ESA Member States, NASA, and Canada. Some of the results in this paper have been derived using the following packages: \texttt{classy} \cite{lesgourgues2011cosmic, class2}, \texttt{healpy} \cite{healpix}, \texttt{Matplotlib} \cite{matplotlib} and \texttt{NumPy} \cite{numpy}. N.B. and S.M. acknowledge partial financial support by ASI Grant No. 2016-24-H.0. A.R.~acknowledges funding from MIUR through the ``Dipartimenti di eccellenza'' project Science of the Universe. We acknowledge partial support from INFN through the InDark initiative.

\bibliographystyle{style_bib}
\bibliography{bibliography}

\end{document}